\documentclass[journal,twoside]{IEEEtran}
\usepackage{cite}
\usepackage{amsmath,amssymb,amsfonts}
\usepackage{algorithmic}
\usepackage{graphicx}
\usepackage{textcomp}

\usepackage{graphicx}
\usepackage{color}
\usepackage[list=true]{subcaption}
\usepackage{enumerate}
\usepackage{multicol}
\usepackage{rotating}
\usepackage{mathtools}
\usepackage{sidecap}
\usepackage{siunitx}
\usepackage{placeins}
\usepackage{comment}
\usepackage{sidecap}

 \usepackage{xcolor}

\usepackage{multirow}
\usepackage{makecell}
\usepackage{paralist}
\usepackage{cancel}
\usepackage{cases} 
\usepackage{color, colortbl}
\usepackage[labelfont=bf,textfont=it,belowskip=0pt,aboveskip=5pt,tableposition=top]{caption}
\usepackage[colorlinks=true,
citecolor=blue,
filecolor=black,
linkcolor=blue,
urlcolor=blue]{hyperref}
\usepackage{longtable}
\usepackage{graphbox}
\usepackage{xspace}

\graphicspath{{figures/}{tables/}}

\newcounter{runidnum}

\newcommand{\secref}[1]{\S\ref{#1}}

\newcolumntype{R}{>{\columncolor{gray!20}}r}
\newcolumntype{L}{>{\columncolor{gray!20}}l}
\definecolor{Gray}{gray}{0.85}
\newcolumntype{a}{>{\columncolor{Gray}}c}

\newcommand{\NM}{\textbf{\textit{NME}}}
\newcommand{\M}{\textbf{\textit{ME}}}

\newcommand{\elltwoT}{\ensuremath{\pi_{\text{TC}}}} 
\newcommand{\elltwoVT}{\ensuremath{\pi_{\text{VT}}}} 

\newcommand{\mpPatientBrain}[1][]{\ifthenelse { \equal {#1} {} }
    { \vect{m}_D }   
    { \vect{m}_{D,#1} } }  

\newcommand{\mpWarpedAtlasBrain}[1][]{\ifthenelse { \equal {#1} {} }
    { \vect{m}_A^{(1,1)} }   
    { \vect{m}_{A,#1}^{(1,1)} }}   

\newcommand{\ns}[1]{\ensuremath{\mathbf{#1}}}

\newcommand{\idiv}{\ensuremath{\mbox{div}\,}}
\newcommand{\igrad}{\ensuremath{\nabla}}

\newcommand{\p} {\partial}

\newcommand{\vect}[1]{\boldsymbol{#1}} 
\newcommand{\mat}[1]{\boldsymbol{#1}}  

\newcommand{\bipa}{\begin{inparaenum}[(\itshape i\upshape)]}
\newcommand{\eipa}{\end{inparaenum}}

\newcommand{\bipasub}{\begin{inparaenum}[(\itshape a\upshape)]}
\newcommand{\eipasub}{\end{inparaenum}}

\newcommand{\ipoint}[1]{\textit{\textbf{\color{darkgray}#1}}}

\newcommand{\defeq}{\ensuremath{\mathrel{\mathop:}=}}

\newcommand{\T}{\ensuremath{\mathsf{T}}}

\definecolor{sred}{cmyk}{0.01,0.98,0,0.2} 
\reversemarginpar
\newcommand{\mmargin}[1]{{\marginpar{\em\tiny #1}}}\renewcommand{\mmargin}[1]{}

\begin{document}
\title{Ensemble inversion for brain tumor growth models with mass effect}
\author{Shashank Subramanian, Klaudius Scheufele, Naveen Himthani, Christos Davatzikos, \IEEEmembership{Fellow, IEEE}, and George Biros, \IEEEmembership{Member, IEEE}
\thanks{S. Subramanian, Oden Institute, University of Texas at Austin, Texas, USA e-mail: shashanksubramanian@{utexas}.edu.}
\thanks{K. Scheufele, Oden Institute, University of Texas at Austin, Texas, USA e-mail: klaudius.scheufele@{gmail}.com}
\thanks{N. Himthani, Oden Institute, University of Texas at Austin, Texas, USA e-mail: naveen@{oden}.utexas.edu}
\thanks{C. Davatzikos, University of Pennsylvania, Philadelphia, USA e-mail: christos.davatzikos@pennmedicine.{upenn}.edu}
\thanks{G. Biros, Oden Institute, University of Texas at Austin, Texas, USA e-mail: biros@{oden}.utexas.edu.}
}
\maketitle

\begin{abstract}
We propose a method for extracting physics-based biomarkers from a \emph{single} multiparametric Magnetic Resonance Imaging (mpMRI) scan bearing a glioma tumor.  We account for mass effect, the deformation of brain parenchyma due to the growing tumor, which on its own is an important radiographic feature but its automatic quantification remains an open problem. In particular, we calibrate a partial differential equation (PDE) tumor growth model that captures mass effect, parameterized by \emph{a single scalar parameter}, tumor \emph{proliferation}, \emph{migration}, while localizing the \emph{tumor initiation site}. 
The single-scan calibration problem is severely ill-posed because the precancerous, healthy, brain anatomy is unknown.  To address the ill-posedness, we introduce an ensemble inversion scheme that uses a number of \emph{normal subject brain templates} as proxies for the healthy precancer subject anatomy. 
We verify our solver on a synthetic dataset and perform a retrospective analysis on a clinical dataset of 216 glioblastoma (GBM) patients. We analyze the reconstructions using our calibrated biophysical model and demonstrate that our solver provides both global and local quantitative measures of tumor biophysics and mass effect. We further highlight the improved performance in model calibration through the inclusion of mass effect in tumor growth models---including mass effect in the model leads to 10\% increase in average dice coefficients for patients with significant mass effect. 
We further evaluate our model by introducing novel biophysics-based features and using them for survival analysis. Our preliminary analysis suggests that including such features can improve patient stratification and survival prediction. 

\end{abstract}


\vspace{5pt}
\begin{small}
{\textit{(Preprint submitted to IEEE Transactions on Medical Imaging)}}
\end{small}
\vspace{-5pt}
\section{Introduction}
\label{sec:intro}

Computational oncology is an emerging field that attempts to integrate biophysical models with imaging with the goal of assisting image analysis in a clinical setting.  Typically, the integration is accomplished by calibrating PDE model parameters from images.  A significant challenge in brain tumors, and in particular, GBMs, is the single-scan calibration, that becomes even harder in the presence of mass effect.   However, modeling provides the capability for automatic quantification of mass effect along with additional biomarkers related to infiltration and tumor aggressiveness.  Mass effect alone, is quite significant as it is an important radiographic marker~\cite{Steed2018,prasannaMassEffectDeformation2019a,Baris:2016}.
Here, we propose a biophysical model calibrated using a single pretreatment scan. Longitudinal pretreatment scans are rare for GBMs since immediate treatment (surgical resection and subsequent radiotherapy) is necessary for most patients.

\subsection{Contributions} 
The single-scan calibration problem is formidable for two main reasons: the tumor initial condition (IC) and the subject's healthy precancerous brain anatomy (the brain anatomy/structural MRI before the cancer begins to grow) are unknown. 
To circumvent this, a template (normal subject brain from another healthy individual) is used as a \emph{proxy} to the healthy subject brain. However, natural anatomical differences between the template and the subject interfere with tumor-related deformations; disentangling the two is hard.  In light of these difficulties, our contributions (as an extension to our recently published work~\cite{Subramanian2020MICCAI}) are as follows:
\textbullet \,  Based on the method described in~\cite{Subramanian20IP}, we propose a novel multistage scheme for inversion: first, we estimate the tumor initial conditions, then, given a template, we invert for scalar model parameters representing tumor proliferation, migration, and mass effect. We repeat this step for several templates and we compute expectations of the observables (see~\secref{sec:methods}).  
\textbullet \, These calculations are quite expensive. However, the entire method runs in parallel on GPUs so that 3D inversion takes less than two hours for any subject. We open-source our parallel solver software on Github\footnote{\url{https://github.com/ShashankSubramanian/GLIA}}.
\textbullet \, We use synthetic data (for which we know the ground truth) to verify our solver and estimate the errors associated with our numerical scheme.
\textbullet \, We apply our methods on a dataset of 216 GBM patients 
and evaluate our model in terms of its ability to match observed tumor patterns given the calibration:  three scalar parameters and the tumor spatial initial condition.
Futher, we introduce a novel feature extraction methodology to obtain biophysics-based features and demonstrate their value in an important clinical task---patient stratification and overall survival prediction. 
We report these results in~\secref{sec:results}. 

\subsection{Related work}
There have been many attempts to integrate PDE models with imaging data~\cite{Mang2020,Lipkova:2019a,mang2012biophysical,Hogea:2008b,Gholami:2016a}.
The most common mathematical models for tumor growth dynamics are based on reaction-diffusion PDEs~\cite{swanson2000quantitative,konukoglu2010image}, which have been coupled with mechanical models to capture mass effect~\cite{hogea2007modeling,hormuth2018}. While there have been many studies to calibrate these models using inverse problems~\cite{konukoglu2010image,Hogea:2008b,Gholami:2016a,Gooya:2012a}, most do not invert for all unknown parameters (tumor initial condition and model parameters) or they assume the presence of multiple imaging scans (both these scenarios make the inverse problem more tractable). In our previous work~\cite{Subramanian20IP}, we presented a methodology to invert for tumor initial condition (IC) and cell proliferation and migration from a single scan; but we did not account for mass effect. We demonstrated the importance of using a \emph{sparse tumor IC} (see \secref{sec:methods}) to correctly reconstruct for other tumor parameters.
In~\cite{Lipkova:2019a}, the authors consider a Bayesian framework for estimating model parameters but do not consider mass effect and use a single seed for tumor initial condition. The absence of mass effect in the tumor models allows for the use of simiplistic precancer brain approximations through deformed templates (computed by deformable registration)~\cite{Lipkova:2019a} or simple tissue replacement strategies~\cite{Scheufele2020}.
Other than those works,  the current state of the art for single-scan biophysically-based tumor characterization is GLISTR~\cite{Gooya:2012a}.

The main shortcoming of GLISTR is that it requires manual seeding for the tumor IC and uses a single template. Further, GLISTR uses deformable registration for both anatomical variations and mass effect deformations but does not decouple them (this is extremely ill-posed), since the primary goal of GLISTR lies in image segmentation. Hence, the estimated model parameters can be unreliable due to the unknown extent to which registration accounts for mass effect and tumor shapes.
Finally, the authors in~\cite{Abler2019} present a 2D synthetic study to quantify mass effect. However, they assume known tumor IC and precancer brain.
To the best of our knowledge, we are not aware of any other framework that can fully-automatically calibrate tumor growth models with mass effect for all unknown parameters in 3D.
Furthermore, our solvers employ efficient parallel algorithms and GPU acceleration which enable realistic solution times, an important consideration for clinical applications.

\section{Methodology}
\label{sec:methods}
We introduce the following notation: $c = c(\vect{x},t)$ is the tumor concentration ($\vect{x}$ is a voxel, $t$ is time) with \emph{observed} tumor data  $c_1 = c(\vect{x}, 1)$ and  \emph{unknown} tumor IC $c_0 = c(\vect{x}, 0)$; $\vect{m}(\vect{x},t) \defeq (m_{\text{WM}}(\vect{x},t), m_{\text{GM}}(\vect{x},t), m_{\text{VT}}(\vect{x},t), m_{\text{CSF}}(\vect{x},t))$ is the brain segmentation into white matter (WM), gray matter (GM), ventricles (VT), and cerebrospinal fluid (CSF)\footnote{We denote CSF to be the non-ventricular cerebrospinal fluid substructure of the brain} with \emph{observed} preoperative segmented brain $\vect{m}_p = \vect{m}(\vect{x}, 1)$ and \emph{unknown} precancer healthy brain $\vect{m}_0 = \vect{m}(\vect{x}, 0)$; Additionally, ($\kappa$, $\rho$, $\gamma$) are \emph{calibration} scalars representing \emph{unknown} migration rate,  proliferation rate, and mass effect intensity.

\subsection{Tumor growth mathematical model}\label{subsec:forward}
Following~\cite{Subramanian:2019a}, we use a non-linear reaction-advection-diffusion PDE (on an Eulerian framework):
\begin{subequations}
    \label{e:fwd:mass}
    \begin{align}
    \p_t c + \idiv (c\vect{v}) - \kappa \mathcal{D} c  - \rho \mathcal{R} c & = 0 ~ \mbox{in}~\Omega \times (0,1] \label{e:reac_diff}  \\
        \p_t \vect{m} + \idiv (\vect{m} \otimes\vect{v}) & = 0 ~\mbox{in}~ \Omega \times (0,1] \label{e:m_transport}  \\
    \idiv (\lambda \igrad \vect{u} + \mu (\igrad \vect{u} + \igrad\vect{u}^\T)) & = \gamma \igrad c ~ \mbox{in}~\Omega \times (0,1] \label{e:linear_elasticity} \\
   \p_t \vect{u} & = \vect{v}\label{e:velocity} ~ \mbox{in}~\Omega \times (0,1],
    \end{align}
\end{subequations}
where $\mathcal{D} :=  \idiv k_{\vect{m}} \igrad c$ is a diffusion operator; $\mathcal{R} := \rho_{\vect{m}} c(1 - c)$ is a logistic growth operator; $k_{\vect{m}}$ and $\rho_{\vect{m}}$ control the spatial heterogeneity of the diffusion and reaction coefficients in different brain tissues. 
Eq.~\eqref{e:reac_diff} is coupled to a linear elasticity equation (Eq.~\eqref{e:linear_elasticity}) with forcing $ \gamma \igrad c$, which is coupled back through a convective term with velocity $\vect{v}(\vect{x},t)$ which parameterizes the displacement $\vect{u}(\vect{x},t)$. 
The linear elasticity
model is parameterized by  Lam\`e coefficients $\lambda(\vect{x,t})$ and $\mu(\vect{x},t)$, whose values depend on the tissue type $\vect{m}(\vect{x},t$).
We note here that $\gamma = 0$ implies $\vect{u} = \vect{0}$, which reduces the tumor growth model to a simple reaction-diffusion PDE with no mass effect (i.e, Eq.~\eqref{e:reac_diff} with $\vect{v} = \vect{0}$).
Solution of Eq.~\eqref{e:fwd:mass} requires initial conditions for the tumor $c_0$ and the precancer brain anatomy $\vect{m}_0$.

Following~\cite{Subramanian20IP}, we parameterize $c_0 =  \boldsymbol{\phi}^T(\vect{x}) \vect{p} = \sum_{i=1}^{m} \phi_i(\vect{x}) p_i$;
where $\vect{p}$ is an $m$-dimensional parameterization vector, $\phi_i(\vect{x}) = \phi_i(\vect{x}-\vect{x_i}, \sigma)$ is a Gaussian function centered at point $\vect{x}_i$ with standard deviation $\sigma$, and $\boldsymbol{\phi}(\vect{x}) = \{\phi_i(\vect{x}) \}_{i = 1}^m$. Here, $\vect{x}_i$ are voxels that are segmented as tumor and $\sigma$ is one voxel, meaning $m$ can be quite large ($\sim$1000). This parameterization alleviates some of the ill-posedness associated with the inverse problem~\cite{Subramanian20IP}.

\subsection{Inverse problem}
The unknowns in our growth model are $(\vect{p},\vect{m}_0,\kappa, \rho, \gamma)$.
We calibrate $\vect{p} \in\ns{R}^{m}, \kappa, \rho, \gamma \in\ns{R}$ and treat $\vect{m}_0$ as a random variable on which we take expectation by solving the calibration problem for multiple $\vect{m}_0$, representing different normal templates (see numerical scheme~\secref{subsec:scheme}). That is, given $\vect{m}_0$, we solve the following inverse problem:
\begin{subequations}\label{eq:tumor_objective}
    \begin{equation}
    \min_{(\vect{p}, \kappa, \rho, \gamma)}
\frac{1}{2} \| \mat{O}c(1) - c_1 \|_2^2 + \\ \frac{1}{2} \| \vect{m}(1) - \vect{m}_p \|_2^2 + \frac{\beta}{2} \| \boldsymbol{\phi}^T \vect{p} \|_2^2
    \end{equation}
    
    \begin{numcases}{\hspace{-80pt}\mbox{s.t.}~}
       \mathcal{F}(\vect{p}, \kappa, \rho, \gamma) &\mbox{given by}~\eqref{e:fwd:mass}, \label{eq:state_constraint} \\
          \| \vect{p} \|_0  \leq s  &\mbox{in}~$\ns{R}^{m}$, \label{eq:sparsity_constraint}\\
          \max (\boldsymbol{\phi}^T \vect{p})  = 1  &\mbox{in}~$\Omega$. \label{eq:max_constraint}
    \end{numcases}
\end{subequations}
The objective function minimizes the $L_2$ mismatch between the simulated tumor $c(1) = c(\vect{x},1)$ at $t = 1$ and preoperative tumor data $c_1$ as well as the $L_2$ mismatch between the deformed precancer brain (due to the tumor) $\vect{m}(1) = \vect{m}(\vect{x},1)$ at $t = 1$ and the preoperative patient brain data $\vect{m}_p$. The mismatch terms are balanced by a regularization term on the inverted initial condition (IC). $\mat{O}$ is an observation operator that defines the clearly observable tumor margin (see~\secref{subsec:select_atlas} for its definition; for additional details see\cite{Scheufele2020}).
Following~\cite{Subramanian20IP}, we introduce two additional constraints to our optimization problem---Eq.~\ref{eq:sparsity_constraint} parameterizes the tumor initial condition using at most $s$ Gaussians and with Eq.~\ref{eq:max_constraint} we assume that at $t = 0$ the tumor concentration reaches one at some voxel in the domain. Both Eq.~\ref{eq:sparsity_constraint} and Eq.~\ref{eq:max_constraint} are \emph{modeling assumptions}. 
We use them to deal with the several ill-posedness of the backward tumor growth PDE. In~\cite{Subramanian20IP}, we discuss this in detail.

\subsection{Summary of our multistage inversion method}\label{subsec:scheme}
To solve Eq.~\ref{eq:tumor_objective}, we propose the following iterative scheme: 
\begin{enumerate}[(\textbf{S}.1)]
  \item First, we estimate $\vect{p}$, ignoring mass effect, using the patient brain anatomy. That is, we solve Eq.~\ref{eq:tumor_objective} using the growth model with no mass effect $\mathcal{F}(\vect{p}, \kappa, \rho, \gamma=0)$ for $(\vect{p},\kappa,\rho)$.\footnote{$\gamma = 0$ indicates no  mass effect and Eq.~\eqref{e:m_transport}, Eq.~\eqref{e:linear_elasticity} and Eq.~\eqref{e:velocity} are not needed} Our \emph{precancer} scan is approximated by $\vect{m}_0 = \vect{m}_p$ with tumor regions replaced by white matter. 
  \item Next, we approximate $\vect{m}_0$ using a normal template scan $\vect{m}_t$.
  Since $\vect{p}$ is defined in $\vect{m}_p$, we register $\vect{m}_p$ to $\vect{m}_t$ and transfer $\vect{p}$ to the $\vect{m}_t$ space\footnote{we simply mask the tumor region for this registration}. This registration is used to ensure that the tumor initial condition does not fall in anatomical structures such as ventricles (where the tumor does not grow).
  \item Finally, we solve Eq.~\ref{eq:tumor_objective} for $\kappa,\rho,$ and $\gamma$ using the estimated $\vect{p}$ and approximation $\vect{m}_0 = \vect{m}_t$.
\end{enumerate}
We repeat (\textbf{S}.2) and (\textbf{S}.3) for an ensemble of templates $\vect{m}_t$ to make our inversion scheme less sensitive to the template selected. 
In (\textbf{S}.1), we employ the fast \emph{adjoint-based} algorithm outlined in~\cite{Subramanian20IP} that alternates between an $\ell_0$ projection to enforce sparsity and an $\ell_2$ regularized solve for the unknowns to obtain a \emph{sparse} and \emph{localized} tumor initial condition.
For the $\ell_2$ regularized solve in (\textbf{S}.1) and the inverse solve in (\textbf{S}.3), we use a quasi-Newton optimization method (L-BFGS) globalized by Armijo linesearch with gradient-based convergence criteria. For (\textbf{S}.3) we employ first order finite differences to approximate the gradient since there are only three scalar unknowns.

\subsection{Selection of template and numerical parameters}\label{subsec:select_atlas}
We observe variability in estimated parameters across different $\vect{m}_t$. We were able to reduce this variability by selecting $\vect{m}_t$ that are more similar to $\vect{m}_p$ using ventricle size as similarity.
Since tumor mass effect typically compresses the ventricles, we select those templates that have larger ventricular volumes than the patient and within a relative margin of 30\%.
This margin (30\%) is computed by simulating several large tumors with strong mass effect and close proximity to the ventricles to estimate the extent of mass-effect-induced ventricle compression.
Here, we assume that the ventricles are a \emph{qualitative indicator} of patient-template anatomical similarity. 
More accurate similarities are problematic to compute as they rely on image registration which is unable to decouple tumor mass effect from anatomical differences. 

Next, we discuss the selection of additional parameters used in our solver. These parameters are default and do not require any subject-specific tuning.
\begin{enumerate}[(i)]
  \item We use 80 normal subject scans from the Alzheimer’s Disease Neuroimaging Initiative (ADNI)\footnote{ADNI database \url{adni.loni.usc.edu}} normal/control subject dataset as our sample space for templates. In order to get good expectation estimates, it is preferable to use a large number of templates. We choose 16 templates for
  two reasons: 
  First, we were unable to consistently\footnote{greater than 50\% of the patient cohort is able to yield sufficient number of candidate templates based on our selection criterion} find a larger number of templates that satisfy our preselection criteria across the 216 GBM patient clinical dataset, and second, the sensitivity of inversion to the number of templates does not alter (drop) significantly with larger number of templates 
  (see discussion on sensitivity to number of templates in appendix Tab. A3).
  \item Experimental evidence suggests that glioma cells infiltrate faster in white matter~\cite{giese1996migration}. We choose $\kappa_{\text{GM}} = 0.2\kappa_{\text{WM}}$, where $\kappa_{\text{GM}}$ and $\kappa_{\text{WM}}$ are the tumor diffusion rates in the gray matter and white matter, respectively. Literature~\cite{swanson2000quantitative,Lipkova:2019a,Gooya:2012a} suggests values of 0.1 and 0.2. We do not observe any significant sensitivity of our inversion to either choice (also observed in the independent study in GLISTR~\cite{Gooya:2012a}).
  \item Elastic properties for different tissue types are fixed based on~\cite{Subramanian:2019a}. See appendix Tab. A1.
  \item Optimizer convergence parameters (tolerance, maximum iterations) are selected based on~\cite{Scheufele2020} using tests which balance run times and solution accuracy on synthetic datasets (see~\secref{sec:results} and appendix Tab. A1).
  \item The standard deviation $\sigma$ of the Gaussian functions is taken as the resolution of our input MRI image. We select Gaussians from an equally-spaced\footnote{The spacing between Gaussians is $\delta = 2\sigma$, which is empirically obtained to guarantee good conditioning of $\int_{\Omega} \boldsymbol{\phi}\boldsymbol{\phi}^T d\vect{x}$} candidate set of Gaussians at voxels that are segmented as tumor. These form our basis functions $\phi_i(\vect{x})$ (see~\secref{subsec:forward}) for the tumor IC $c_0$.
  \item The sparsity level $s$ of the tumor IC is based on an empirical study on a large clinical cohort~\cite{Scheufele2020}---the study observed values between 3 and 6 to show a good compromise between a localized sparse solution and the reconstruction of complex tumor shapes (as measured by the dice coefficient). We select $s$ to be 5. Note that $s$ is an upper bound on expected number of activation sites for the tumor IC; the actual number of sites is determined by solving Eq.~\ref{eq:tumor_objective}.
  \item We define our observation operator application as $\mat{O}c(\vect{x},1) := o(\vect{x}) \odot c(\vect{x},1)$ (see~\eqref{eq:tumor_objective}). Here, $\odot$ signifies hadamard product and $o(\vect{x})$ is defined to be one in every voxel $\vect{x}$ except the peritumoral edema (ED), where it is zero. That is, we observe the tumor concentration $c$ everywhere except within the edema where it is set (observed) as zero. We make this choice because the peritumoral region in GBMs is known to be infiltrated with cancer cells to an \emph{unknown} extent~\cite{Lemee:2015:peritumoral}. The observation operator can be interpreted as complete uncertainty of tumor observation in this region. Hence, the tumor profile in the peritumoral edema is governed solely by the model dynamics, and not fitted to arbitrary values (see~\cite{Scheufele2020} for more details).
\end{enumerate}

\subsection{Biophysics-based feature extraction}\label{subsec:feature_extraction}
Here we introduce a set of manually crafted features that are solely based on the output of the biophysical model and can potentially be used in downstream clinical tasks. 
\begin{enumerate}[1.]
    \item \ipoint{Spatial extraction: } We extract features related to tumor concentration, displacement, and stresses at various times of the tumor evolution, including \emph{future} times. We compute the following spatial fields: $c$, $\partial c/ \partial t$, $\|\igrad{c}\|_2$, $\| \vect{u} \|_2$, $\text{trace}(\mat{T})$, $\tau$, $\mathcal{J} := \text{det}(\mat{I} + \igrad \vect{u})$, $\| \vect{v} \|_2$; where $\mat{T} := \lambda\idiv{\vect{u}}\mat{I} + \mu(\igrad{\vect{u}} + \igrad{\vect{u}^T})$ is the elasticity stress tensor, $\tau$ is the maximum shear stress field\footnote{$\tau$ can be computed as $1/2(e_3-e_1)$, with $e_1$ and $e_3$ as the minimum and maximum eigenvalues of the stress tensor $\mat{T}$}, $\text{det}$ is the determinant, and $\mathcal{J}$ is the deformation Jacobian. Then, given any of these scalar fields $y(\vect{x})$, we compute the following features: $\int_\Omega y d\Omega$, standard deviation (quantifies spread), $87.5^{\text{th}}$, $95^{\text{th}}$, and $99^{\text{th}}$ percentiles (quantifies magnitudes), and surface integral $\int_\Gamma y d\Gamma$ at the isosurface $c = c^\star$ for prespecified $c^\star$, which we approximate by:  
\begin{equation}
\int_\Gamma y(\vect{x}) d\Gamma = \frac{N^{1/3}}{2\pi} \int_\Omega \mathbf{1}(|c(\vect{x}) - c^\star| < \epsilon) \odot y(\vect{x}) d\vect{x}
\end{equation}
for some tolerance $\epsilon$\footnote{$\epsilon$ is computed experimentally as $0.1$ to ensure a continuous but sufficiently narrow surface}. Here, $N$ is total number of voxels, $\mathbf{1}$ is an indicator function, and $\odot$ indicates hadamard product. To compute $c^\star$, we first calculate the segmentation of the brain---each voxel is assigned the label (healthy tissues or tumor) with highest concentration value. Then, we compute $c^\star = c(\vect{x}^\star)$, where $\vect{x}^\star = \text{argmax}_{\vect{x} \notin \text{tumor}}(\|\igrad{c}\|_2)$; this is the isosurface at the tumor front where the tumor shows strong gradients and represents the evolving front. 
  The volumetric features are computed for different spatial regions: healthy brain $\mathcal{B}$ (that is, every non-tumorous brain voxel), the tumor region, and the full spatial domain $\Omega$.
In total, we compute $128$ features at each time $t$.
    \item \ipoint{Temporal extraction: }These features are computed at 5 different time points: diagnosis/imaging scan time and the time instances at which volume fraction of the tumor is 0.1\%, 1\%, 3\%, and 5\% of the brain. We have chosen these times to \emph{temporally standardize} all our features, since every patient (at diagnosis) might be at a different stage of cancer evolution. For example, for patients with large tumors, the features are mostly computed at past times,
  and for patients with smaller tumors, they represent forecasts (assuming no treatment). The biophysical model makes such forecasts possible.
\end{enumerate}
In~\secref{subsec:clinical}, we demonstrate the utility of the above extracted features in an important clinical task.

\subsection{Workflow}
\label{subsec:workflow}
Here we summarize the workflow we used to process a clinical subject dataset.
\subsubsection{Image preprocessing}\label{subsec:preproc}
Our dataset consists of single-time-point (preoperative) structural mpMRI scans (T1, T1-CE, FLAIR, T2) of 216 GBM patients, along with age, gender, surgical resection status, and overall survival information.
For each patient, the four mpMRI scans are first affinely registered to a template atlas. 
Then, they are segmented into tumor regions of enhancing tumor (ET), necrotic tumor (NEC), and peritumoral edema (ED) using a state-of-the-art neural network~\cite{myronenko2018} trained on the BraTS~\cite{brats:2018} dataset which provides expert radiologist ground truth labels for these tumor regions. 
Since our tumor model is a single-species growth model, we define the tumor region/label as the tumor core (TC) $:= \text{ET} \cup \text{NEC}$.
Finally, they are further segmented into healthy substructures (GM, WM, VT, and CSF) using ANTs~\cite{avants2009}, where
we register several templates to any given patient and ensemble the atlas-based segmentation for the different healthy tissue labels.
The segmentation labels serve as a proxy for our tissue (and tumor) concentration data. 
\subsubsection{Inversion software}
Our open-sourced software is written in C++ and parallelized using MPI and CUDA. The solver is capable of running on several CPUs as well as GPUs (a single GPU suffices for medical imaging resolutions). A forward solve of our mass effect model on $256^3$ spatial dimensions takes less than 2 minutes on a V100 NVIDIA GPU and the full inversion takes an average of 1--2 hours (depending on the input tumor data and number of $\vect{m}_t$ templates in the ensembled inversion). We refer the reader to~\cite{Subramanian20IP,Gholami:2017a} for our solver numerics and parallel algorithms.

\section{Results}
\label{sec:results}

\subsection{Synthetic verification}\label{subsec:synthetic}
In~\cite{Subramanian2020MICCAI}, we verified the correctness of the solver using synthetic data.
We briefly summarize the tests and results here.
We ask the following three questions:
\begin{enumerate}[\bf(Q1)\upshape] \item Given $\vect{p}$ (tumor IC) and $\vect{m}_0$ (precancer scan), can we reconstruct $(\kappa, \rho, \gamma)$ (i.e., the diffusion, reaction, and mass effect coefficients) using (\textbf{S}.3) from~\secref{subsec:scheme}? \item Given $\vect{m}_0$ but unknown $\vect{p}$, can we reconstruct $(\vect{p}, \kappa, \rho, \gamma)$ using (\textbf{S}.1) and (\textbf{S}.3)? \item With $\vect{m}_0$ and $\vect{p}$ unknown, can we reconstruct $(\vect{p}, \kappa, \rho, \gamma)$ using (\textbf{S}.1)-(\textbf{S}.3), using ensembled inversion with $\vect{m}_0 = \vect{m}_t$? \end{enumerate} 
To create the synthetic data, we select $\vect{m}_0, \vect{p}, \kappa, \rho,$ and $\gamma$ and run our forward solver to $t = 1$. The model outputs at $t = 1$ define our synthetic data which is then used to reconstruct $(\vect{p}, \kappa, \rho, \gamma)$, after which we compare the errors.
We consider the following test-cases (variations) for our parameter configurations:

\medskip
\begingroup
\setlength{\tabcolsep}{2pt}
\begin{tabular}{@{}ccc}
    \textit{TC(a):} & no mass effect & $(\gamma^\star, \rho^{\star}, \kappa^{\star}) = (0, 12, 0.025)$   \\
    \textit{TC(b):} & mild mass effect & $(\gamma^\star, \rho^{\star}, \kappa^{\star}) = (0.4, 12, 0.025)$   \\
    \textit{TC(c):} & medium mass effect & $(\gamma^\star, \rho^{\star}, \kappa^{\star}) = (0.8, 10, 0.05)$   \\
    \textit{TC(d):} & large mass effect & $(\gamma^\star, \rho^{\star}, \kappa^{\star}) = (1.2, 10, 0.025)$   \\
\end{tabular}
\endgroup \\

\noindent where $^\star$ represents the non-dimensionalized \emph{ground truth} parameters. 
The tumors along with the deformed $\vect{m}_0$ anatomy are visualized in appendix Fig. A1.

We report relative errors for $\kappa,\rho,$ and $\gamma$. We compute displacement two-norm field $u = u(\vect{x}) = \|\vect{u}(\vect{x})\|_2$ (this field quantifies mass effect deformation) and report the relative error in the norm $\|u\|_2$ and the infinity norm $u_\infty = \|u\|_\infty$ (in mm).
Finally, we report the dice coefficient in reconstructing the tumor core $\elltwoT$.

 \medskip \noindent
\textbf{(Q1)} \textit{Known $\vect{m}_0$, known $\vect{p}$: }
We reproduce our inversion results in appendex Tab. A2.
We observe excellent reconstruction for all test-cases with relative errors less than 2\%.

\medskip \noindent
\textbf{(Q2)} \textit{Known $\vect{m}_0$, unknown $\vect{p}$: }We use our inversion scheme outlined in~\secref{subsec:scheme}, where we first invert for the tumor initial condition using (\textbf{S}.1) and then the model parameters with \emph{known} precancer scan. We reproduce our inversion results in appendix Tab. A2 and observe that the model parameters are still recovered well (see appendix for discussion).

\begin{figure}[!htbp]
    \centering
    \includegraphics[width=0.5\textwidth]{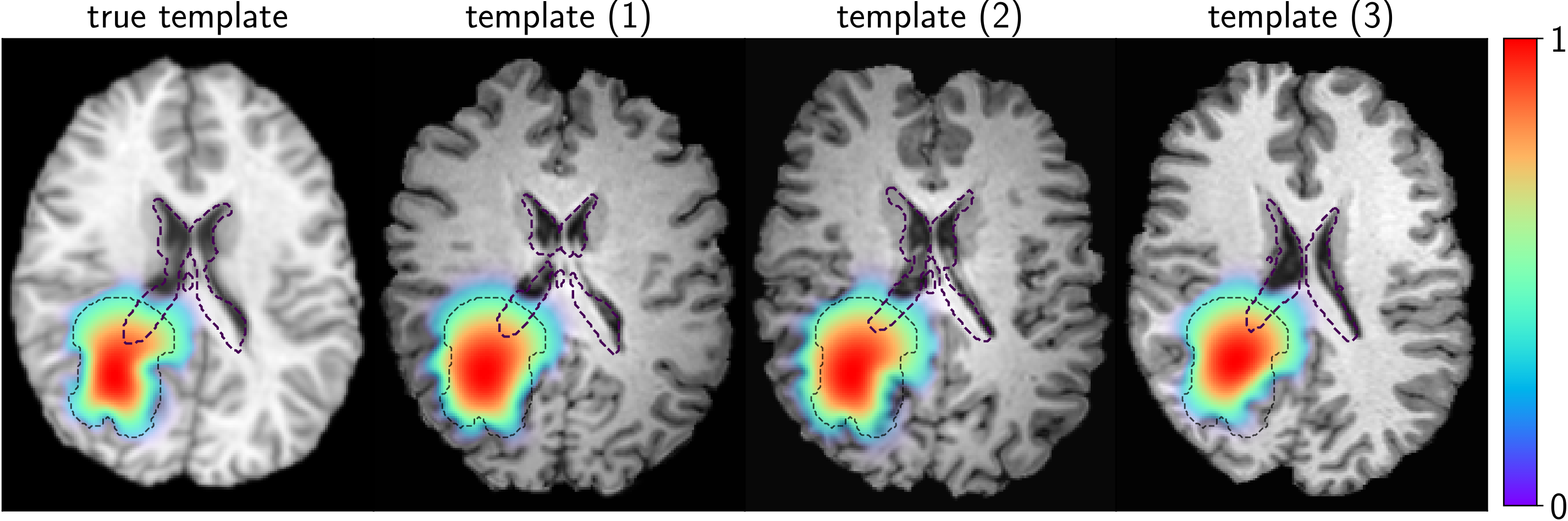}
    \caption{Reconstruction of the patient using 16 templates (of which three templates are shown here) as the precancer scan. The tumor data segmentation is highlighted as a black dashed contour line. The reconstructed tumor concentration is shown in color. The undeformed ventricle configuration is highlighted as a purple dashed line to indicate the extent of mass effect in the reconstructions.} \label{fig:syn-atlas-recon}
\end{figure}

\medskip \noindent
\textbf{(Q3)} \textit{Unknown $\vect{m}_0$, unknown $\vect{p}$: }This scenario corresponds to the actual clinical problem. 
We invoke (\textbf{S}.2) and use an ensemble of 16 templates to approximate $\vect{m}_0$ (see~\secref{subsec:select_atlas}). 
We report inversion results in Tab.~\ref{tab:syn-unknown-c0-unknown-hp} and show an exemplary reconstruction of the patient using the different templates in Fig.~\ref{fig:syn-atlas-recon}.
We report the expectations (mean; $\mu_\iota$) and standard deviations ($\sigma_\iota$; characterizes the sensitivity of our inversion to each template) for each parameter $\iota = (\gamma, \rho, \kappa, u_\infty)$. We refer to the tumor-deformed template ($\vect{m}(\vect{x},1)$) with the grown tumor ($c(\vect{x},1)$) using the reconstructed parameters as the \emph{reconstructed patient}.

For the tumor dice coefficient, we compute the probability $p_l(\vect{x})$ for any tissue/tumor label $l$ at location $\vect{x}$ as:
    \begin{equation}\label{eq:prob_atlas}
      p_l(\vect{x}) = \frac{\sum_{t=1}^M\mathbf{1}(s_t(\vect{x}) = l)}{M},
    \end{equation}
where $s_t(\vect{x})$ is the label map, $M$ is the total number of templates ($16$ in our case), and $\mathbf{1}$ is an indicator function. Then, we construct a \emph{probabilistic label image} by simply assigning every voxel to the label with maximum probability. The tumor dice is computed using the tumor label from the probabilistic reconstructed patient and the patient tumor data. 

Despite not knowing the true $\vect{m}_0$, we are still able to capture the parameters.  The expectation of the displacement norm error has an average error of about 17\%, while the expectations of reaction and diffusion coefficients show about 24\% and 15\% average relative errors, respectively. The reaction coefficient shows least sensitivity (standard deviations are an order of magnitude smaller than the expectations); the diffusion coefficient also shows small sensitivity which is roughly the order of numerical diffusion that is introduced in our forward solvers\footnote{We employ spectral solvers that require some artificial diffusion through smoothing to prevent aliasing errors}; the maximum displacement shows standard deviations of the order of $1$ mm. The mass effect parameter $\gamma$ shows high standard deviations for the test-cases with small mass effect and progressively improves as the mass effect increases. 
This is due to ill-conditioning with respect to $\gamma$  when either the tumor or the mass effect are small.

\setlength{\tabcolsep}{6pt}
\begin{table*}
    \caption{Inversion results for \textbf{unknown} $\vect{m}_0$ and \textbf{unknown} $\vect{p}$ using the multistage inversion scheme (see~\secref{subsec:scheme}). 
    We report the ground truth parameters $\iota^\star$ and the mean ($\mu_\iota$) and standard deviation ($\sigma_\iota$) of the reconstructed parameters across the ensemble of templates; $\iota = \gamma, \rho, \kappa,$ and the maximum value of the displacement field two-norm (i.e., $\|u\|_\infty$ in mm). 
    $e_\iota$ is the relative error for the {average} value of biophysical parameter $\iota = \gamma, \rho, \kappa$; $e_u$ is the relative error in the {average} norm of the displacement field two-norm (i.e., $\|u\|_2$), and $\elltwoT$ is the probabilistic tumor core dice coefficient.
    (Note: when the ground truth ($^\star$) is zero, the errors are absolute errors; diffusion coefficient ($\kappa$) values are scaled by $\num{1E-2}$ for clarity).
    \label{tab:syn-unknown-c0-unknown-hp}
    }
    \centering
            \begin{scriptsize}
    \makebox[\textwidth]{\centering
        \begin{tabular}{c|ca|ca|ca|ca|a}
            \hline
            Test-case & $(\gamma^\star, \mu_\gamma, \sigma_\gamma)$ & $e_\gamma$ &$(\rho^\star, \mu_\rho, \sigma_\rho)$ & $e_\rho$ &$(\kappa^\star, \mu_\kappa, \sigma_\kappa)/0.01$ & $e_\kappa$ &$(u_\infty^\star, \mu_{u_\infty}, \sigma_{u_\infty})$ & $e_u$ & \elltwoT \\
            \hline    
            \textit{TC (a)} & (0,0.19,0.15) & \num{1.97E-1} & (12,8.96,0.65) & \num{2.53E-1} & (2.5,1.89,0.74) & \num{2.41E-1} & (0,2.1,1.51) & \num{1.76E1} & \num{7.93E-1} \\
            \textit{TC (b)} & (0.4,0.43,0.2) & \num{7.93E-2} & (12,8.82,0.32) & \num{2.65E-1} & (2.5,2.41,0.61) & \num{3.34E-2} & (5.4,4.93,2.21) & \num{8.48E-2} & \num{8.2E-1} \\
            \textit{TC (c)} & (0.8,0.69,0.2) & \num{1.25E-1} & (10,8,0.15) & \num{2E-1} & (5,4.23,0.83) & \num{2.41E-1} & (8.98,6.89,1.86) & \num{2.22E-1} & \num{8.48E-1} \\
            \textit{TC (d)} & (1.2,1.1,0.14) & \num{7.76E-2} & (10,7.78,0.26) & \num{2.23E-1} & (2.5,2.69,0.72) & \num{7.67E-2} & (13.35,10.43,1.09) & \num{2.02E-1} & \num{8.39E-1} \\
           \hline 
    \end{tabular}}
           \end{scriptsize}
\end{table*}

\subsection{Clinical data analysis}\label{subsec:clinical}
We apply our inversion scheme on the 216 clinical GBM patient cohort (see~\secref{subsec:workflow} for details regarding the dataset and preprocessing steps). We report exemplary reconstruction results in Tab.~\ref{tab:clinical} for four patients with different ranges of mass effect. We observe similar trends in sensitivity (standard deviations) of parameters and dice coefficients as in the synthetic test-cases. 

\begin{figure}[!htbp]
    \centering
    \includegraphics[width=0.5\textwidth]{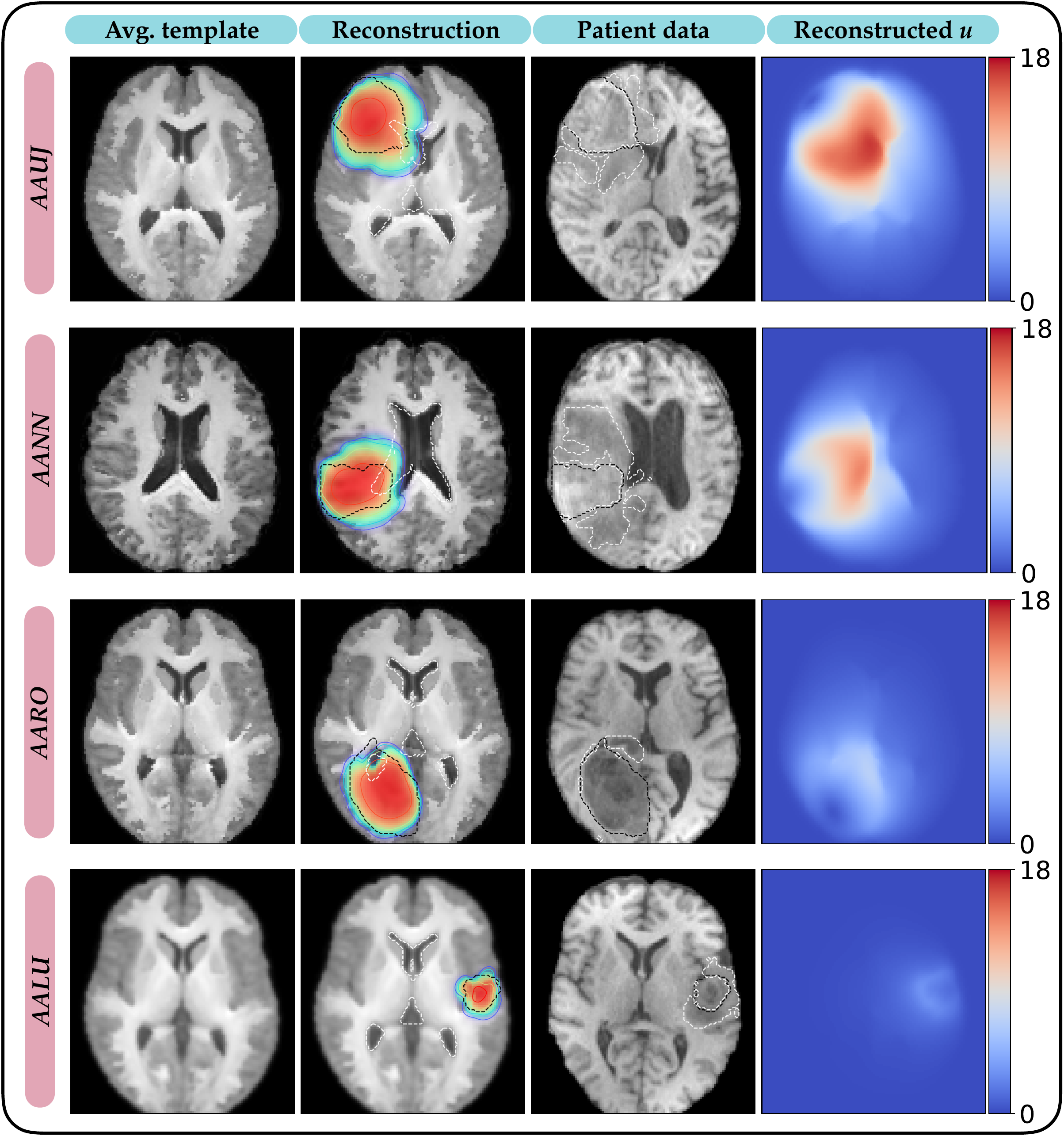}
    \caption{
      Reconstructed images from the inversion scheme. We show four representative subjects (IDs in parenthesis). The first column shows the averaged template T1 MRI, the second column depicts the averaged patient reconstruction with the estimated tumor concentration (higher concentrations ($\sim$1) are indicated in red and lower ones in green) overlayed. We also show the undeformed ventricle contour (white dashed line) to express the extent of mass effect, as well as the TC data contour (black dashed line). The third column shows the patient T1 MRI data with the TC data contour (black dashed line) and the edema data contour (white dashed line). The last column shows the reconstructed average displacement (in mm) two-norm field.
      }\label{fig:clinical}
\end{figure}

\begin{figure*}[!htbp]
    \centering
    \includegraphics[width=\textwidth]{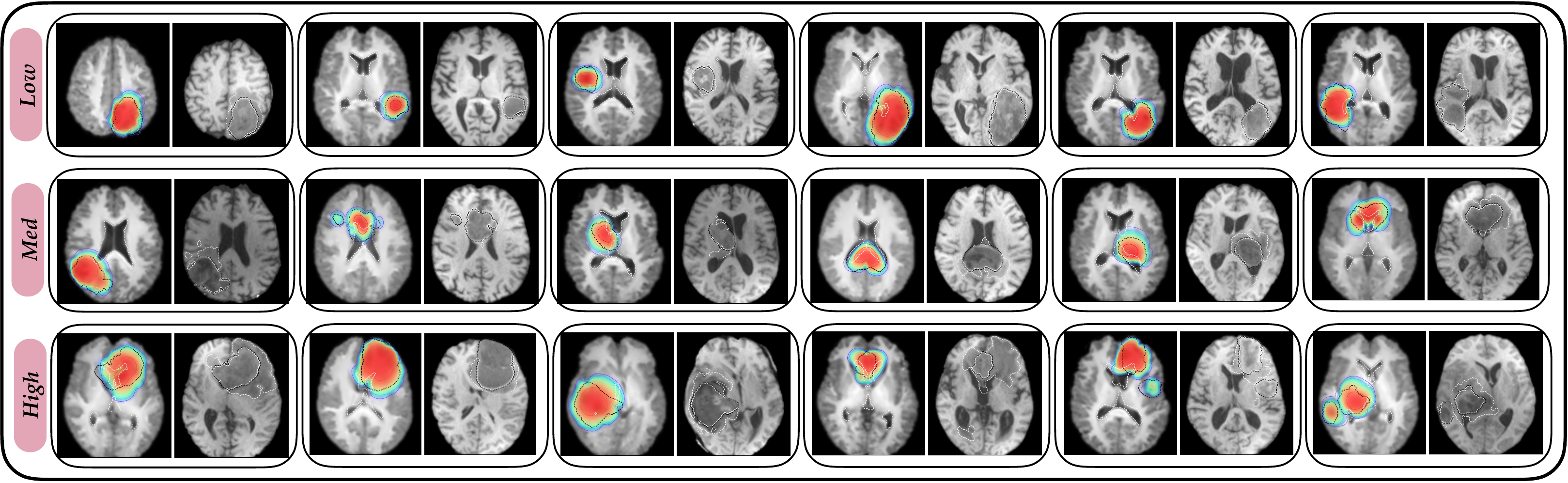}
    \caption{Some clinical reconstructions stratified into low, medium, and high mass effect based on equal quantiles of displacement. Each image is a pair indicating the reconstructed patient with estimated tumor concentrations (left) and patient T1 MRI data (right). The TC data (black dashed line) and undeformed ventricles (white dashed line) are visualized in the reconstructed patient (left) and TC data (black dashed line) and edema data (white dashed line) are visualized in the patient data (right).
    }\label{fig:clinical_many}
\end{figure*}

We visualize the reconstructions in Fig.~\ref{fig:clinical}. We show the averaged\footnote{Averaging results in coarse representations of the ensemble and is, hence, only used for visualization purposes} probabilistic template image, the averaged reconstructed patient image, the patient data, and the average reconstructed displacement two-norm field. We observe good quantitative agreement of the tumor (dice coefficients greater than 83\%) and good quantitative characterization of the mass effect. Patients \emph{AAUJ} and \emph{AANN} present large tumors with significant mass effect ($\|u\|_\infty \sim 18\text{mm}\pm1$ and $12\text{mm}\pm5$, respectively); patient \emph{AARO} also shows a large tumor but with mild mass effect ($\|u\|_\infty \sim 7\text{mm}\pm1$); patient \emph{AALU} presents a small tumor with negligible mass effect($\|u\|_\infty \sim 2\text{mm}\pm1$). In Fig.~\ref{fig:clinical_many}, we show results stratified according to the extent of mass effect. 

\setlength{\tabcolsep}{5pt}
\begin{table*}
    \caption{Inversion results for clinical data. 
 We report the mean ($\mu_\iota$) and standard deviation ($\sigma_\iota$) of the reconstructed parameters across the ensemble of templates; $\iota = \gamma, \rho, \kappa,$ and the maximum value of the displacement field two-norm (i.e., $u_\infty = \|u\|_\infty$ in mm). 
     We also report the probabilistic tumor core dice coefficient $\elltwoT$, the average runtime $\mu_T$, and standard deviation in runtime $\sigma_T$ in (hh:mm:ss).
    (Note: the diffusion coefficient ($\kappa$) values are scaled by $\num{1E-2}$ for clarity).
    \label{tab:clinical}
    }
    \centering
            \begin{scriptsize}
    \makebox[\textwidth]{\centering
        \begin{tabular}{c|cccc|a|a}
            \hline
            Patient ID & $(\mu_\gamma, \sigma_\gamma)$ & $(\mu_\rho, \sigma_\rho)$ & $(\mu_\kappa, \sigma_\kappa)/0.01$ & $(\mu_{u_\infty}, \sigma_{u_\infty})$ & $\elltwoT$ & $(\mu_T, \sigma_T)$ (in hh:mm:ss) \\

            \hline    
          \textit{AAUJ} & (0.82, 0.081) & (9.92, 0.75) & (1.87,0.89) & (17.57, 1.4) & 0.875 & (01:13:27,00:13:43)\\
          \textit{AANN} & (0.59, 0.25) & (9.93, 0.61) & (2.28, 1.23) & (11.68, 4.91) & 0.853 & (01:27:24,00:06:23) \\
          \textit{AARO} & (0.4, 0.089) & (10.64, 0.58) & (1.75, 0.38) & (7.23, 1.49) & 0.839 & (01:08:19,00:05:47) \\
          \textit{AALU} & (0.36, 0.23) & (6.39, 0.52) & (1.1, 0.55) & (2.43, 1.48) & 0.942 & (01:05:22,00:08:14) \\
           \hline 
    \end{tabular}}
           \end{scriptsize}
\end{table*}
In order to understand the role and importance of mass effect, we conduct a preliminary analysis on the patient cohort using two different growth models: model {\M} represents a growth model with mass effect and model {\NM} represents a model without mass effect (i.e., $\gamma = 0$ or a simple nonlinear reaction-diffusion PDE).
We compare the two using the three questions below:
\begin{enumerate}[\bf(CQ1)\upshape] 
\item Does model {\M} improve the reconstruction dice coefficients?
\item Is mass effect (as quantified through model {\M}) simply a feature of larger tumors?
\item Do features extracted from our biophysical model (see~\secref{subsec:feature_extraction}) add value to important clinical evaluations? 
\end{enumerate} 

\begin{figure*}[!htbp]
    \centering
	  \subcaptionbox{\label{fig:dice}}{\includegraphics[width=0.65\textwidth]{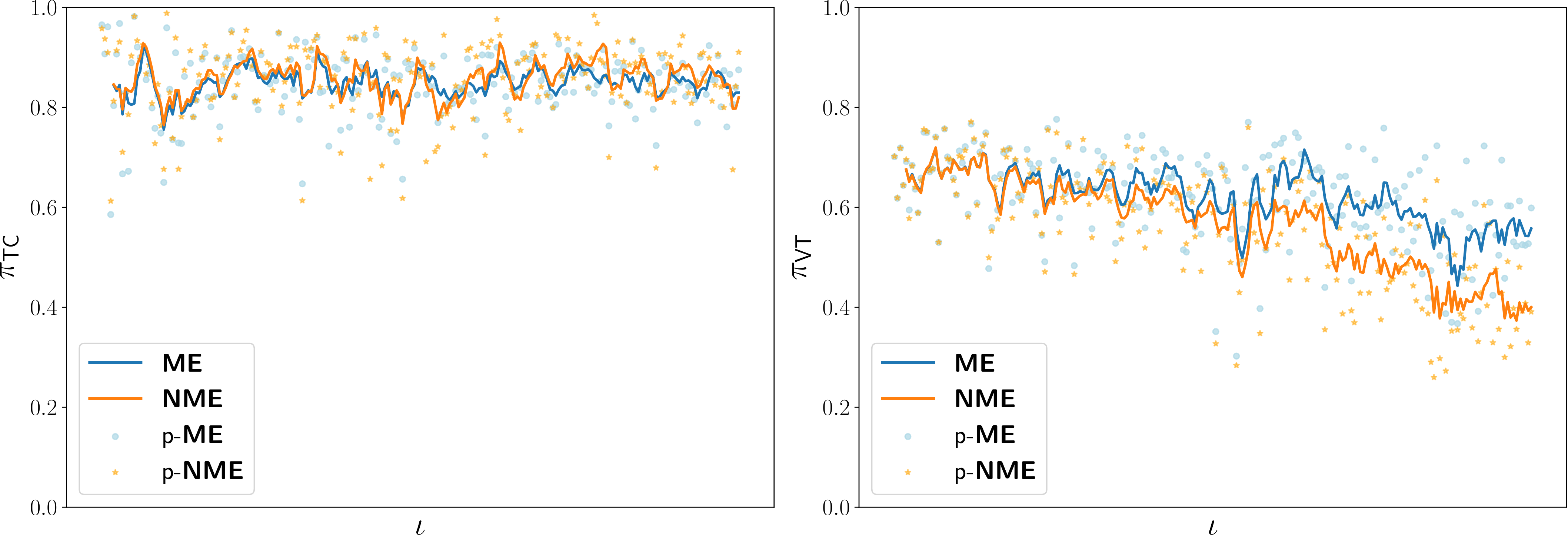}}%
	  \hfill
	  \subcaptionbox{\label{fig:vol}}{\includegraphics[width=0.29\textwidth]{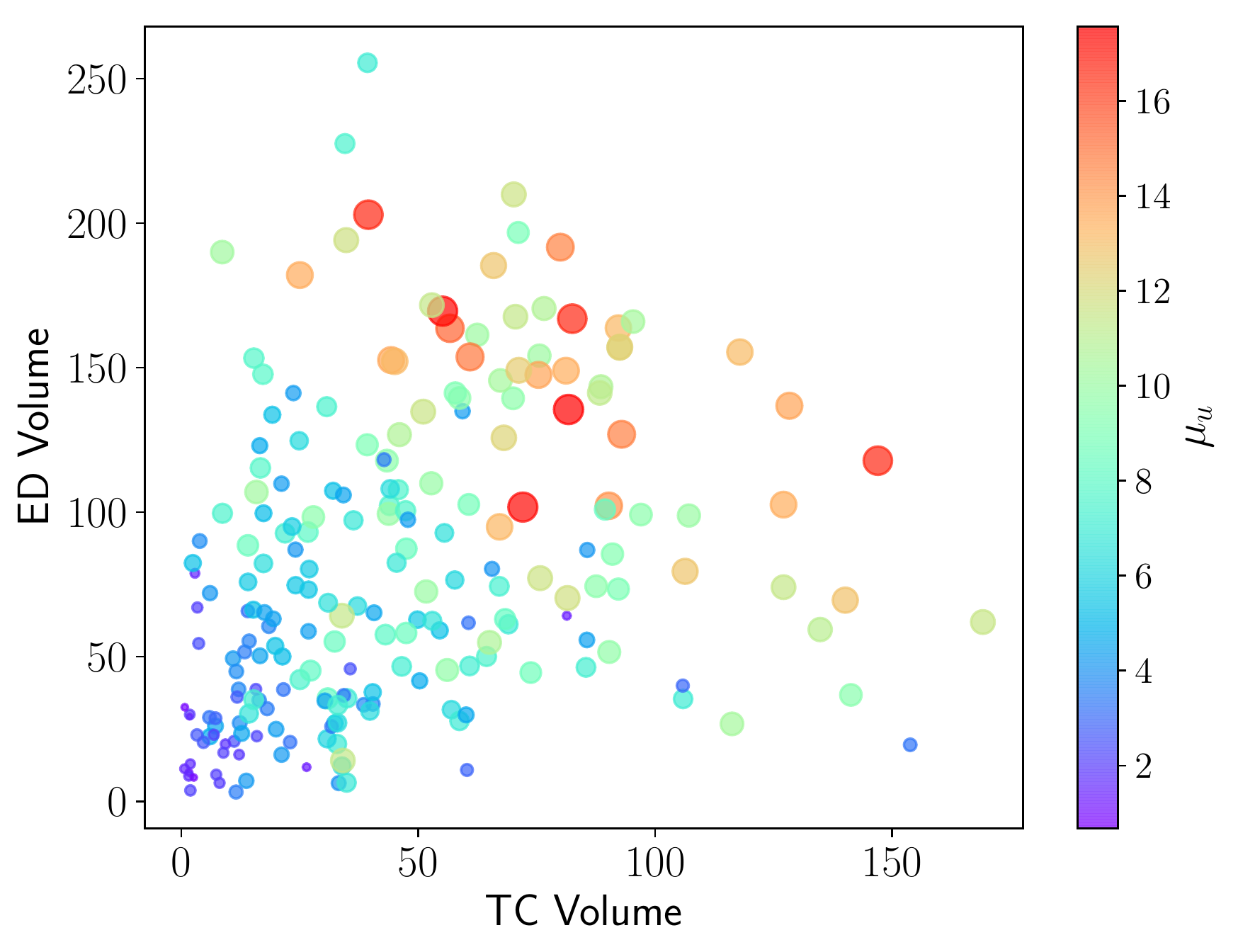}}%
    \caption{
      (a) \textbf{CQ1}: Plots of tumor core (TC, left) dice coefficient $\elltwoT$ and ventricle (VT, right) dice coefficient $\elltwoVT$ vs patient ID index $\iota$ sorted according to increasing mass effect for the two models {\M} and {\NM}. The lines ({\M},{\NM}) represent a rolling 5-patient mean and $p$-$\M$ and $p$-$\NM$ are the raw data points. We observe similar reconstruction for TC, but significant performance gains are visible in the VT reconstruction.
      (b) \textbf{CQ2}: 
  Scatter plot of tumor core (TC) volume vs peritumoral edema (ED) volume (volumes in $cm^3$). Each data point is sized and colored according to the mean displacement field $\mu_u$---larger (and red) circles are indicative of patients with high mass effect. We observe variability in mass effect deformations across tumors with similar volume. For example, tumors with TC volume approximately 75 $cm^3$ show a wide range of mass effect.
      }
    \label{fig:CQ12}
\end{figure*}

\medskip \noindent
\textbf{(CQ1)} We plot the dice coefficients (see~\eqref{eq:prob_atlas}) for the tumor core (TC) and ventricles (VT) in Fig~\ref{fig:dice}. We sort the results in increasing order of mass effect (using the reconstructed $u_\infty$ and $\|u\|_2$); the hypothesis is that {\M} is more informative as mass effect becomes a pronounced feature of the disease. We observe that the tumor dice coefficients ($\elltwoT$) show similar performance for both {\M} and {\NM}. The average TC dice coefficients across the population are 0.849 and 0.857 for {\M} and {\NM}, respectively.  This is not surprising since {\NM} selects $\vect{p}, \kappa,$ and $\rho$ to match the observed tumor core and does not consider matching the anatomy. In contrast, {\M} has to account for both the anatomy and the tumor. For the VT dice, we observe that {\M} shows an overall better performance that progressively improves as the mass effect increases. Across the population, the average VT dice coefficients are 0.621 and 0.563 for {\M} and {\NM}, respectively---an approximate 6\% performance improvement. For patients whose displacement norms are greater than the median displacement norms (median is 6.76 mm; we consider these patients to qualitatively have moderate to large mass effect), this improvement is approximately 10\%.

Here, we note that the VT dice coefficients are low due to the significant anatomical variability between the templates and the patient. However, this can be remedied with a subsequent image registration between the reconstructed patient and the patient data---the deformation from mass effect has been \emph{decoupled} through our inversion scheme and, now, registration can be used to correct for anatomical differences.

\begin{figure}[!htbp]
    \centering
    \includegraphics[width=0.4\textwidth]{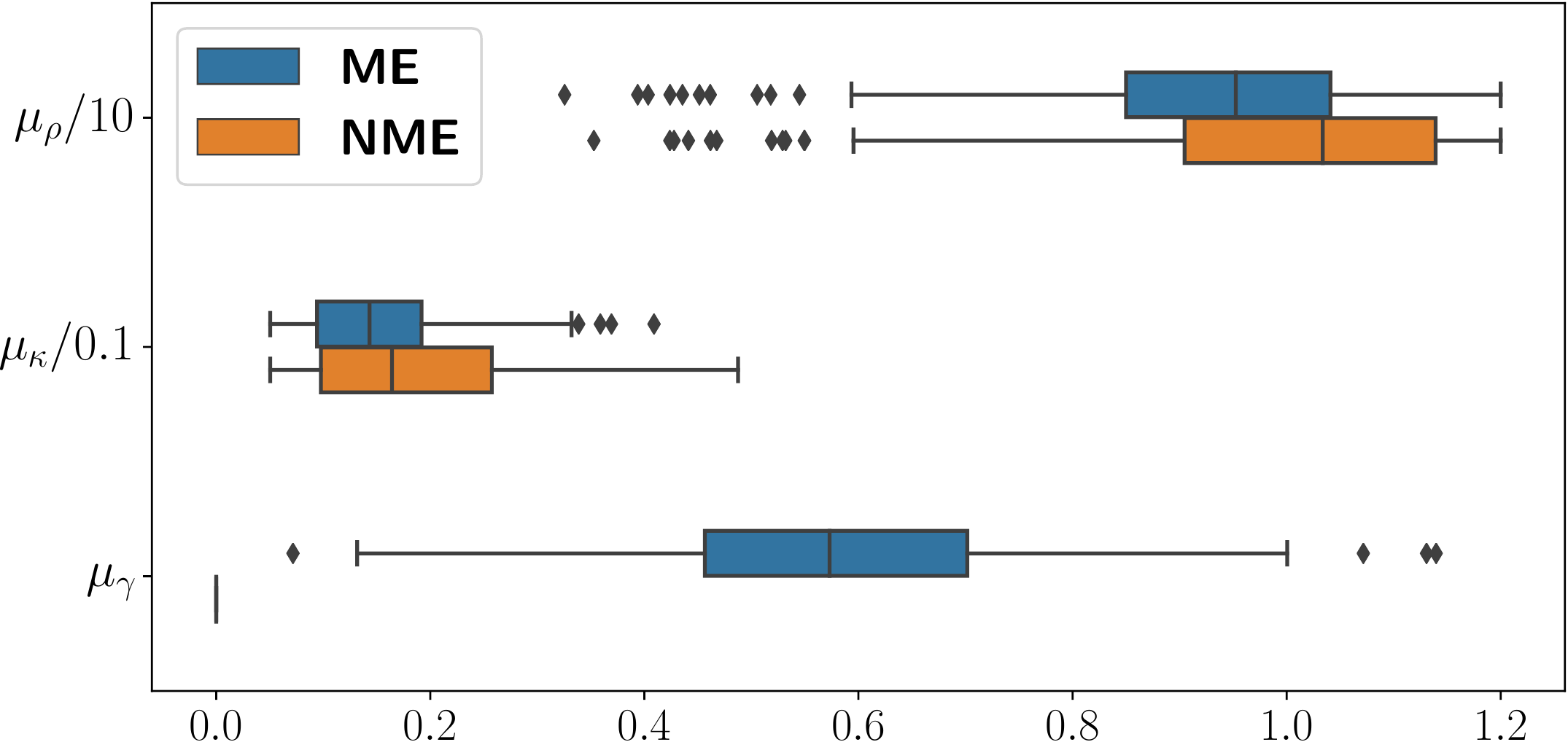}
    \caption{Box plots of the mean reaction, diffusion, and mass effect coefficients ($\mu_{\rho}, \mu_{\kappa}, \mu_{\gamma}$, respectively) for the patient cohort using the two models {\M} and {\NM}. We observe mass effect in the range $\mu_u = 7.18 \pm 3.93$ mm ($\mu_u$ is the mean of $u_\infty = \|u\|_\infty$, where $u$ is the displacement two-norm field).}\label{fig:param-var}
\end{figure}

We also plot the variability of the biophysical scalar parameters in Fig~\ref{fig:param-var}.  We observe that both models provide a \emph{different} set of reaction and diffusion coefficients. This is because {\NM} captures the tumor but ignores mass effect, which leads to larger $\kappa$ and $\rho$. 
In contrast, {\M} accounts for both growth and anatomy deformation. Interestingly, the fitted parameters exhibit smaller cross-subject variability compared to {\NM}.

\begin{figure*}[!htbp]
    \centering
	  \subcaptionbox{\label{fig:M1-vs-M2-a}}{\includegraphics[width=0.475\textwidth]{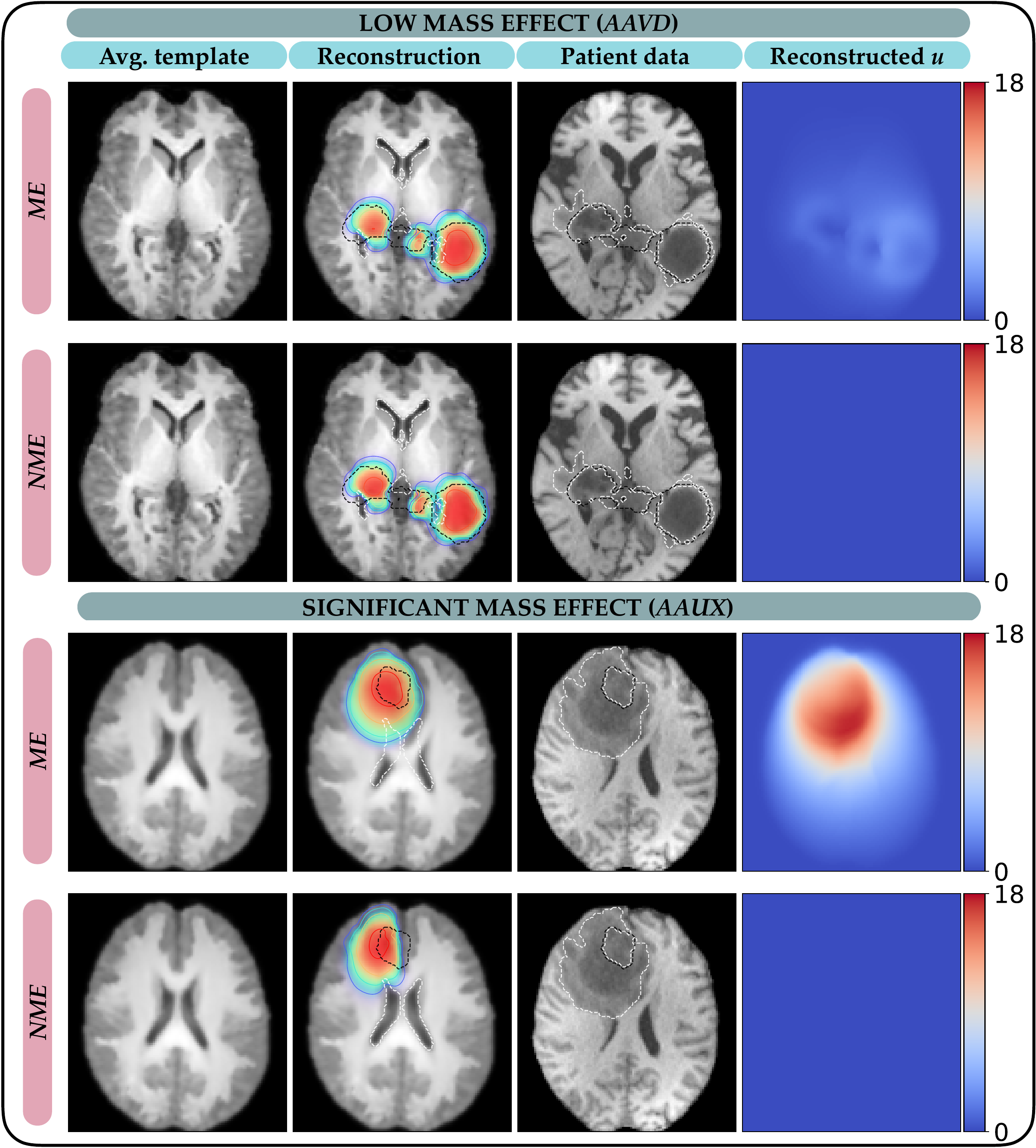}}%
	  \hfill
	  \subcaptionbox{\label{fig:M1-vs-M2-b}}{\includegraphics[width=0.493\textwidth]{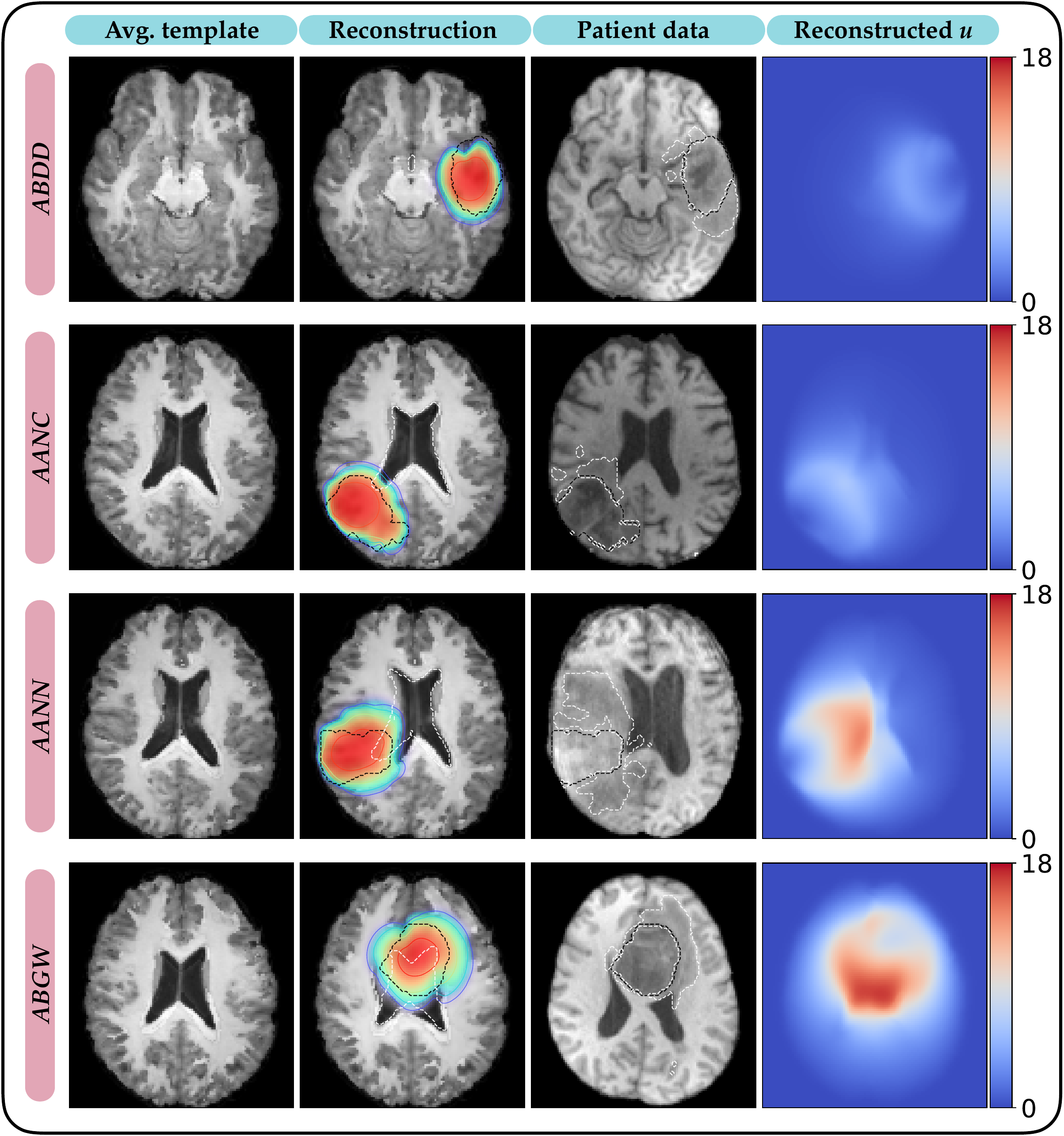}}%
    \caption{Reconstructed images from the inversion scheme. \textbf{(a)} We compare the two models {\M} and {\NM}. We compare the results for one subject with low and one subject with high mass effect. The first column shows the averaged template T1 MRI, the second column shows the averaged patient reconstruction with the estimated tumor concentration (higher concentrations ($\sim$1) are indicated in red and lower ones in green) overlayyed. We also show the undeformed ventricle contour (white dashed line) to express the extent of mass effect, as well as the TC data contour (black dashed line). The third column shows the patient T1 MRI data with the TC data contour (black dashed line) and the edema data contour (white dashed line). The last column depicts the reconstructed average displacement (in mm) two-norm field. 
    Both models exhibit similar reconstruction performance for the tumor and healthy tissues in patient AAVD (low mass effect); for patient AAUX (high mass effect) {\M} outperforms {\NM} because it is capable of deforming the ventricles (and other structures; for example, the model captures the mid-line shift necessary to account for the tumor). 
    \textbf{(b)} Reconstructions for patients with similar TC volume but different ranges of mass effect. Each row represents a subject (with subject ID indicated on the left).  
    We observe different mass effect ranges ($\mu_u = 4.1, 6.9, 11.7, 17.3$ mm from top to bottom) for similar TC volumes ($65.6, 69.0, 70.1, 72.0$ $\text{cm}^3$ from top to bottom). While the first patient shows almost no mass effect, the last patient exhibits significant compression of ventricles (also seen in the deformation from the initial ventricle contours and the displacement field).}\label{fig:M1-vs-M2}
\end{figure*}

In Fig~\ref{fig:M1-vs-M2-a}, we compare the two models. Subject \emph{AAVD} is a multicentric GBM with negligible mass effect; both models show similar dice scores ((TC, VT) dice scores of (0.795, 0.613) and (0.813, 0.641) for {\M} and {\NM}, respectively); subject \emph{AAUX} has significant mass effect (as seen by the compressed ventricles and mid-line shift) and {\M} outperforms {\NM} in its reconstructions ((TC, VT) dice scores of (0.84, 0.527) and (0.842, 0.329) for {\M} and {\NM}, respectively), with nearly 20\% improvement in VT dice coefficients and similar TC dice coefficients. We remark that this improvement is achieved by only one additional scalar parameter ($\gamma$) in our biophysical model.


\medskip \noindent
\textbf{(CQ2)} In Fig.~\ref{fig:vol} we report the patient tumor (TC) and edema (ED) volumes; point color and size correspond to the mean displacement field $\mu_u$. We observe that tumors with similar volume can exhibit a wide range of mass effect. For instance, tumors with volume around 70--80 $\text{cm}^3$ show deformations in the negligible range as well as substantial mass effect ($> 15$ mm). In Fig~\ref{fig:M1-vs-M2-b}, we show tumors with volumes within 10\% of each other, but mass effect displacements ranging from 4 mm to 17 mm (also qualitatively visible in the reconstructed displacement two-norm fields). Thus, we conclude that mass effect is not a feature of larger tumors and can possibly be an additional and valuable biomarker for GBMs. 

\begin{figure}[]
    \centering
    \includegraphics[width=0.5\textwidth]{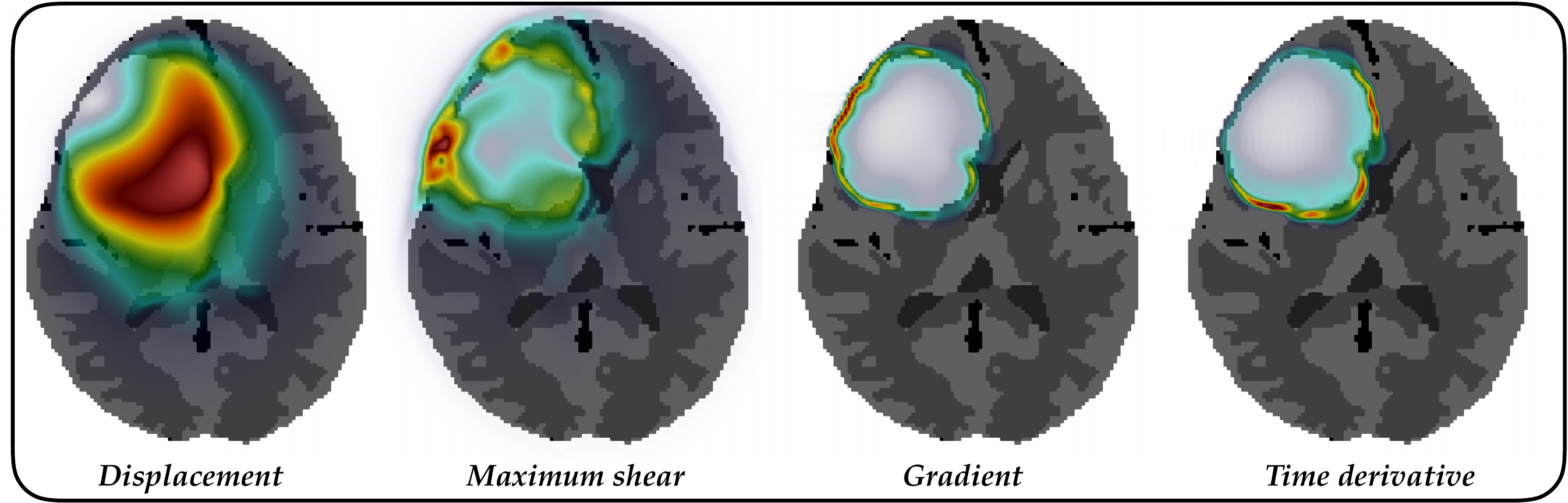}
    \caption{Voxelwise biophysics-based features for a representative patient. We plot 2D slices of $\|\vect{u}\|_2, \tau, \|\igrad{c}\|_2, \partial c/\partial t$ at time instant when the tumor volume fraction is 5\% of the brain volume. All feature values are normalized here (high: red, low: blue)}\label{fig:features_aauj}
\end{figure}

\medskip \noindent
\textbf{(CQ3)} We consider survival data for patients that underwent a gross total surgical resection (GTR)---our survival dataset consists of 141 patients (GTR only) and each patient is classified as a short, mid, or long survivor based on equal quantiles of survival time (in days). 
Our objective is to use patient-specific features to predict the survival class by training a machine learning classifier on our dataset. We consider the following features: \textbullet \, \textit{Age and volumetric features: } Patient age and tumor volumes for each tumor subclass, and \textbullet \, \textit{Biophysics-based features: } Features extracted using our calibrated growth model (see~\secref{subsec:feature_extraction}; we show an example visualization of a few biophysics-based features in Fig~\ref{fig:features_aauj}).
\begin{enumerate}[(i)]
  \item \textit{Feature correlations: }We plot the Spearman correlation of statistically significant feature correlations with survival time in Fig.~\ref{fig:spearman}. We observe that age is the maximally correlated feature with survival---an observation consistent with several previous studies~\cite{Menze:2020a,McKinley:2020a}. We also note that biophysics-based features correlate better than volumetric features (tumor volumes; correlations are insignificant and small), and features computed at standardized times (at specific tumor volume fractions) correlate better than features computed at diagnosis/imaging scan time. 
\begin{figure}[!htbp]
    \centering
    \includegraphics[width=0.5\textwidth]{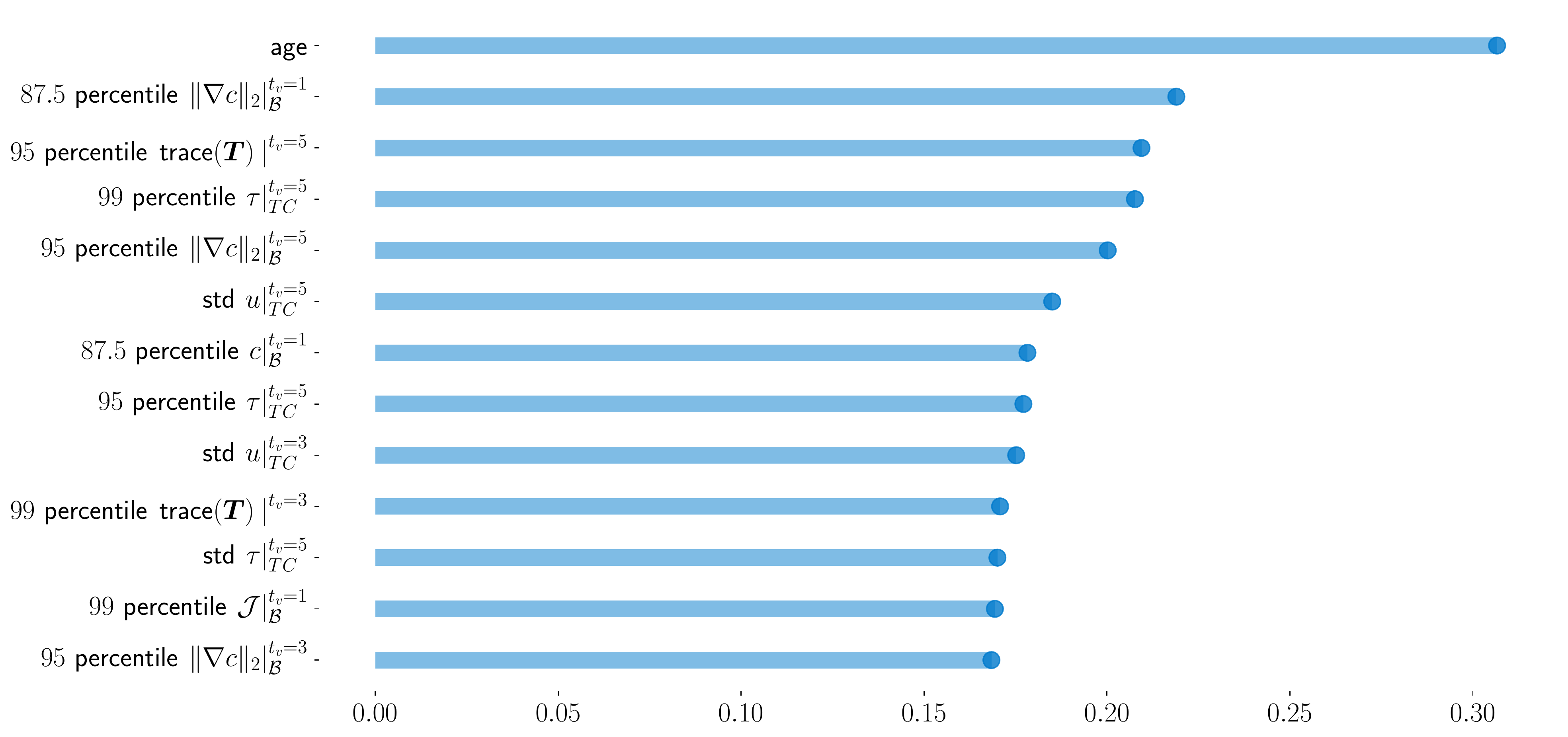}
    \caption{Spearman correlation of features (only statistically significant correlations, $p < 0.05$) with survival sorted in descending order of values. For any feature statistic $y$, $y|_{\mathcal{R}}^{tv=v}$ represents the statistic computed in region $\mathcal{R} \in \{\mathcal{B}, TC\}$ (absent $\mathcal{R}$ denotes no region and statistic is computed using all $\vect{x}$) at time when tumor volume fraction $t_v$ is $v$.}\label{fig:spearman}
\end{figure}

\item \textit{Univariate patient stratification: }We stratify patients based on the median values of a few top correlated features (with survival) and plot the Kaplan-Meier (KM) survival curves for each feature (see Fig.~\ref{fig:KM}). We observe that several biophysics-based features (along with age) dichotomize the patients based on their overall survival. For instance, patients with large values of $\text{trace}(\mat{T})$ (relates to the tumor-induced pressure due to mass effect) show reduced overall survival. We also additionally show the stratification based on a volumetric feature (tumor core volume). We observe volumetric features are unable to stratify the subject survival curves.
\begin{figure}[!htbp]
    \centering
    \includegraphics[width=0.5\textwidth]{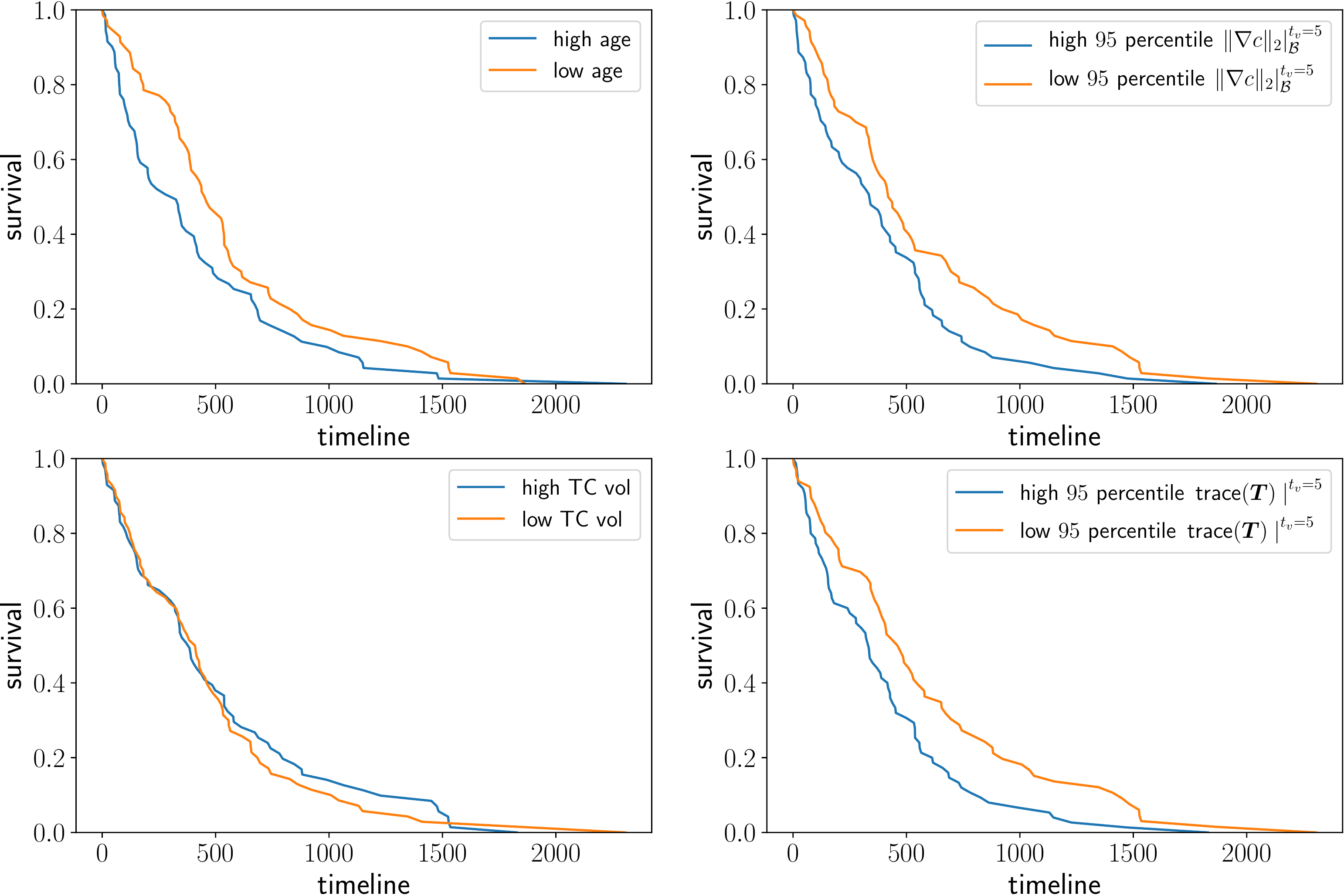}
    \caption{Kaplan-Meier survival curves using median feature value patient stratification for some top correlated (with survival) features and tumor core volume. Age and biophysics feature survival curves are significant (log rank $p < 0.05$). For tumor core volume, log rank $p = 0.72$ and is insignificant---subjects are not stratified based on volumetric features.}\label{fig:KM}
\end{figure}

  \item \textit{Survival classification: }We employ random forests
  as our survival classification model on an 80-20 train-test data split.
We perform a univariate feature selection where we first sort features according to their average 5-fold cross-validated (CV) random forest feature importance. Then, we compute a cross-validated (CV) training accuracy and standard deviation (std) by adding features incrementally in descending order of importances (see appendix Fig. A4). We then select the top 13 features (these include age, edema volume, and other biophysics-based features) since they show maximum CV accuracy and small accuracy standard deviations across the splits (hence prone to lesser overfitting). We report our classification scores in Tab.~\ref{tab:classification}. We observe an improvement in classification scores using the selected features in comparison to using only age and volumetric features. The classifier also shows smaller variability across cross-validated splits and hence, potentially, generalizes well (as suggested in the testing accuracies).
\end{enumerate}

\setlength\tabcolsep{5pt}
\begin{table}
      \caption{Classification of patient survival using age, volumetric, and biophysics-based features. We show the improvement in accuracy by using the features subselected by random forest feature importance---these features include age, volumetric, and biophysics-based features.} \label{tab:classification}
    \centering
            \begin{scriptsize}
    \centering
        \begin{tabular}{c|c|c}
            \hline
            Features & CV training accuracy & Testing accuracy \\
            \hline
            \textit{Age}      &  0.375 $\pm$ 0.107 & 0.276 \\
            \textit{Age + Volumetric} & 0.347 $\pm$ 0.073 & 0.345 \\
            \textit{Selected features} & 0.509 $\pm$ 0.042 & 0.448 \\
            \hline
    \end{tabular}
            \end{scriptsize}
\end{table}

\section{Conclusions}
\label{sec:conclusions}
Our results are very promising. First, our solver is robust (never crashes, takes excessive iterations, or needs subject-specific hyperparameter settings). It  does not require any manual preprocessing and can be run in a black box fashion. Our experiments on  synthetic data with known ground truth is the first time that such a solver is verified and we demonstrated that our approximations do not introduce significant errors despite the fact that ${\vect{m}}_0$ is unknown. For clinical data, the  model errors are expected to dominate the errors from using a template and splitting the calibration procedure into two stages.
Second, we tested our method on a large clinical patient cohort (216 patients) in order to test the feasibility of our method. With a small number of calibration parameters, our solver is able to quantitatively match the observed tumor margins and qualitatively correlate with observed mass effect. Hence, our solver provides a means to quantify and localize mass effect without relying on any assumptions on symmetry and location of the tumor. 
Further, we showed that including mechanical effects in our tumor growth models can provide additional information in terms of capturing the deformations of the tumor and surrounding anatomical structures and results in quantitatively different parameters and reconstructions than a non-biomechanical growth model.
Finally, we introduced a novel biophysics-based feature extraction method and demonstrated the clinical value of the extracted features on an important clinical task---patient stratification and survival prediction using pretreatment imaging scans.
While age remains an important population-based feature for predicting survival outcome, our observations indicate that biophysics-based features can enhance the quantification of survival prognosis. Specifically, these features closely follow the predictive capability of patient age in the following: first, correlation with patient overall survival, second, dichotomization (stratification) of patients based on overall survival, and last, improving prediction of patient overall survival through machine learning methods. 
However, we do note the low survival classification accuracies, indicative of the difficulty in using machine learning methods on small datasets (141 GTR patients here). Similar observations have been made before on the BraTS dataset for survival prediction~\cite{brats:2018}.
Our next step is to incorporate multiple tumor species in our growth model and continue our clinical evaluation using larger patient cohorts. Our hope is that such models can also assist in additional clinical tasks such as tumor recurrence prediction.


\bibliographystyle{bib/IEEEtran}
\bibliography{bib/literature.bib}

\begin{thebibliography}{10}
\providecommand{\url}[1]{#1}
\csname url@samestyle\endcsname
\providecommand{\newblock}{\relax}
\providecommand{\bibinfo}[2]{#2}
\providecommand{\BIBentrySTDinterwordspacing}{\spaceskip=0pt\relax}
\providecommand{\BIBentryALTinterwordstretchfactor}{4}
\providecommand{\BIBentryALTinterwordspacing}{\spaceskip=\fontdimen2\font plus
\BIBentryALTinterwordstretchfactor\fontdimen3\font minus
  \fontdimen4\font\relax}
\providecommand{\BIBforeignlanguage}[2]{{%
\expandafter\ifx\csname l@#1\endcsname\relax
\typeout{** WARNING: IEEEtran.bst: No hyphenation pattern has been}%
\typeout{** loaded for the language `#1'. Using the pattern for}%
\typeout{** the default language instead.}%
\else
\language=\csname l@#1\endcsname
\fi
#2}}
\providecommand{\BIBdecl}{\relax}
\BIBdecl

\bibitem{Steed2018}
T.~C. Steed, J.~M. Treiber, M.~G. Brandel, K.~S. Patel, A.~M. Dale, B.~S.
  Carter, and C.~C. Chen, ``{Quantification of glioblastoma mass effect by
  lateral ventricle displacement},'' \emph{Scientific Reports}, vol.~8, no.~1,
  2018.

\bibitem{prasannaMassEffectDeformation2019a}
P.~Prasanna, J.~Mitra, N.~Beig, A.~Nayate, J.~Patel, S.~Ghose, R.~Thawani,
  S.~Partovi, A.~Madabhushi, and P.~Tiwari, ``\BIBforeignlanguage{en}{Mass
  {{Effect Deformation Heterogeneity}} ({{MEDH}}) on {{Gadolinium}}-contrast
  {{T1}}-weighted {{MRI}} is associated with decreased survival in patients
  with right cerebral hemisphere {{Glioblastoma}}: {{A}} feasibility study},''
  \emph{\BIBforeignlanguage{en}{Scientific Reports}}, vol.~9, no.~1, pp. 1--13,
  Feb. 2019.

\bibitem{Baris:2016}
\BIBentryALTinterwordspacing
M.~M. Baris, A.~O. Celik, N.~S. Gezer, and E.~Ada, ``Role of mass effect, tumor
  volume and peritumoral edema volume in the differential diagnosis of primary
  brain tumor and metastasis,'' \emph{Clinical Neurology and Neurosurgery},
  vol. 148, pp. 67--71, 2016. [Online]. Available:
  \url{https://www.sciencedirect.com/science/article/pii/S0303846716302475}
\BIBentrySTDinterwordspacing

\bibitem{Subramanian2020MICCAI}
S.~Subramanian, K.~Scheufele, N.~Himthani, and G.~Biros, ``Multiatlas
  calibration of biophysical brain tumor growth models with mass effect,'' in
  \emph{Medical Image Computing and Computer Assisted Intervention -- MICCAI
  2020}, A.~L. Martel, P.~Abolmaesumi, D.~Stoyanov, D.~Mateus, M.~A. Zuluaga,
  S.~K. Zhou, D.~Racoceanu, and L.~Joskowicz, Eds.\hskip 1em plus 0.5em minus
  0.4em\relax Cham: Springer International Publishing, 2020, pp. 551--560.

\bibitem{Subramanian20IP}
S.~{Subramanian}, K.~{Scheufele}, M.~{Mehl}, and G.~{Biros}, ``Where did the
  tumor start? an inverse solver with sparse localization for tumor growth
  models,'' \emph{Inverse Problems}, no.~36, 2020.

\bibitem{Mang2020}
\BIBentryALTinterwordspacing
A.~Mang, S.~Bakas, S.~Subramanian, C.~Davatzikos, and G.~Biros, ``Integrated
  biophysical modeling and image analysis: Application to neuro-oncology,''
  \emph{Annual Review of Biomedical Engineering}, vol.~22, no.~1, pp. 309--341,
  2020, pMID: 32501772. [Online]. Available:
  \url{https://doi.org/10.1146/annurev-bioeng-062117-121105}
\BIBentrySTDinterwordspacing

\bibitem{Lipkova:2019a}
J.~Lipkova, P.~Angelikopoulos, S.~Wu, E.~Alberts, B.~Wiestler, C.~Diehl,
  C.~Preibisch, T.~Pya, S.~Comps, P.~Hadjidoukas, K.~V. Leemput,
  P.~Koumoutsakos, J.~Lowengrub, and B.~Menze, ``Personalized radiotherapy
  planning for glioma using multimodal bayesian model calibration,'' \emph{IEEE
  Transactions on Medical Imaging}, vol.~38, no.~8, pp. 1875--1884, 2019.

\bibitem{mang2012biophysical}
A.~Mang, A.~Toma, T.~A. Schuetz, S.~Becker, T.~Eckey, C.~Mohr, D.~Petersen, and
  T.~M. Buzug, ``Biophysical modeling of brain tumor progression: from
  unconditionally stable explicit time integration to an inverse problem with
  parabolic {PDE} constraints for model calibration,'' \emph{Medical Physics},
  vol.~39, no.~7, pp. 4444--4459, 2012.

\bibitem{Hogea:2008b}
C.~Hogea, C.~Davatzikos, and G.~Biros, ``An image-driven parameter estimation
  problem for a reaction-diffusion glioma growth model with mass effect,''
  \emph{J Math Biol}, vol.~56, pp. 793--825, 2008.

\bibitem{Gholami:2016a}
A.~Gholami, A.~Mang, and G.~Biros, ``An inverse problem formulation for
  parameter estimation of a reaction-diffusion model of low grade gliomas,''
  \emph{Journal of Mathematical Biology}, vol.~72, no.~1, pp. 409--433, 2016.

\bibitem{swanson2000quantitative}
K.~Swanson, E.~Alvord, and J.~Murray, ``A quantitative model for differential
  motility of gliomas in grey and white matter,'' \emph{Cell Proliferation},
  vol.~33, no.~5, pp. 317--330, 2000.

\bibitem{konukoglu2010image}
E.~Konukoglu, O.~Clatz, B.~H. Menze, B.~Stieltjes, M.-A. Weber, E.~Mandonnet,
  H.~Delingette, and N.~Ayache, ``Image guided personalization of
  reaction-diffusion type tumor growth models using modified anisotropic
  {Eikonal} equations,'' \emph{Medical Imaging, IEEE Transactions on}, vol.~29,
  no.~1, pp. 77--95, 2010.

\bibitem{hogea2007modeling}
C.~Hogea, C.~Davatzikos, and G.~Biros, ``Modeling glioma growth and mass effect
  in {3D MR} images of the brain,'' in \emph{Medical Image Computing and
  Computer-Assisted Intervention--MICCAI 2007}.\hskip 1em plus 0.5em minus
  0.4em\relax Springer, 2007, pp. 642--650.

\bibitem{hormuth2018}
D.~A. Hormuth, S.~L. Eldridge, J.~A. Weis, M.~I. Miga, and T.~E. Yankeelov,
  ``Mechanically coupled reaction-diffusion model to predict glioma growth:
  Methodological details,'' pp. 225--241, 2018.

\bibitem{Gooya:2012a}
A.~Gooya, K.~M. Pohl, M.~Bilello, L.~Cirillo, G.~Biros, E.~R. Melhem, and
  C.~Davatzikos, ``{GLISTR}: {G}lioma image segmentation and registration,''
  \emph{Medical Imaging, IEEE Transactions on}, vol.~31, no.~10, pp.
  1941--1954, 2013.

\bibitem{Scheufele2020}
K.~{Scheufele}, S.~{Subramanian}, and G.~{Biros}, ``Fully automatic calibration
  of tumor-growth models using a single mpmri scan,'' \emph{IEEE Transactions
  on Medical Imaging}, pp. 1--1, 2020.

\bibitem{Abler2019}
D.~Abler, P.~B{\"u}chler, and R.~C. Rockne, ``Towards model-based
  characterization of biomechanical tumor growth phenotypes,'' in
  \emph{Mathematical and Computational Oncology}, G.~Bebis, T.~Benos, K.~Chen,
  K.~Jahn, and E.~Lima, Eds.\hskip 1em plus 0.5em minus 0.4em\relax Cham:
  Springer International Publishing, 2019, pp. 75--86.

\bibitem{Subramanian:2019a}
S.~Subramanian, A.~Gholami, and G.~Biros, ``Simulation of glioblastoma growth
  using a {3D} multispecies tumor model with mass effect,'' \emph{Journal of
  Mathematical Biology}, vol.~79, no.~3, pp. 941--967, 2019.

\bibitem{giese1996migration}
A.~Giese, L.~Kluwe, B.~Laube, H.~Meissner, M.~E. Berens, and M.~Westphal,
  ``Migration of human glioma cells on myelin,'' \emph{Neurosurgery}, vol.~38,
  no.~4, pp. 755--764, 1996.

\bibitem{Lemee:2015:peritumoral}
J.-M. Lem{\'e}e, A.~Clavreul, and P.~Menei, ``Intratumoral heterogeneity in
  glioblastoma: don't forget the peritumoral brain zone,''
  \emph{Neuro-oncology}, vol.~17, no.~10, pp. 1322--1332, 2015.

\bibitem{myronenko2018}
A.~Myronenko, ``3d mri brain tumor segmentation using autoencoder
  regularization,'' 2018.

\bibitem{brats:2018}
\BIBentryALTinterwordspacing
S.~Bakas, M.~Reyes, and et~al., ``Identifying the best machine learning
  algorithms for brain tumor segmentation, progression assessment, and overall
  survival prediction in the {BRATS} challenge,'' \emph{CoRR}, vol.
  abs/1811.02629, 2018. [Online]. Available:
  \url{http://arxiv.org/abs/1811.02629}
\BIBentrySTDinterwordspacing

\bibitem{avants2009}
B.~B. Avants, N.~Tustison, and G.~Song, ``Advanced normalization tools
  (ants),'' \emph{Insight j}, vol.~2, no. 365, pp. 1--35, 2009.

\bibitem{Gholami:2017a}
A.~Gholami, A.~Mang, K.~Scheufele, C.~Davatzikos, M.~Mehl, and G.~Biros, ``A
  framework for scalable biophysics-based image analysis,'' in \emph{Proc
  ACM/IEEE Conference on Supercomputing}, no.~19, 2017, pp. 19:1--19:13.

\bibitem{Menze:2020a}
F.~Kofler, J.~C. Paetzold, I.~Ezhov, S.~Shit, D.~Krahulec, J.~S. Kirschke,
  C.~Zimmer, B.~Wiestler, and B.~H. Menze, ``A baseline for predicting
  glioblastoma patient survival time with classical statistical models and
  primitive features ignoring image information,'' in \emph{Brainlesion:
  Glioma, Multiple Sclerosis, Stroke and Traumatic Brain Injuries}, A.~Crimi
  and S.~Bakas, Eds.\hskip 1em plus 0.5em minus 0.4em\relax Cham: Springer
  International Publishing, 2020, pp. 254--261.

\bibitem{McKinley:2020a}
R.~McKinley, M.~Rebsamen, K.~Daetwyler, R.~Meier, P.~Radojewski, and R.~Wiest,
  ``Uncertainty-driven refinement of tumor-core segmentation using 3d-to-2d
  networks with label uncertainty,'' 2020.

\end{thebibliography}


\begin{thebibliography}{1}
\providecommand{\url}[1]{#1}
\csname url@samestyle\endcsname
\providecommand{\newblock}{\relax}
\providecommand{\bibinfo}[2]{#2}
\providecommand{\BIBentrySTDinterwordspacing}{\spaceskip=0pt\relax}
\providecommand{\BIBentryALTinterwordstretchfactor}{4}
\providecommand{\BIBentryALTinterwordspacing}{\spaceskip=\fontdimen2\font plus
\BIBentryALTinterwordstretchfactor\fontdimen3\font minus
  \fontdimen4\font\relax}
\providecommand{\BIBforeignlanguage}[2]{{%
\expandafter\ifx\csname l@#1\endcsname\relax
\typeout{** WARNING: IEEEtran.bst: No hyphenation pattern has been}%
\typeout{** loaded for the language `#1'. Using the pattern for}%
\typeout{** the default language instead.}%
\else
\language=\csname l@#1\endcsname
\fi
#2}}
\providecommand{\BIBdecl}{\relax}
\BIBdecl

\bibitem{Subramanian:2019a}
S.~Subramanian, A.~Gholami, and G.~Biros, ``Simulation of glioblastoma growth
  using a {3D} multispecies tumor model with mass effect,'' \emph{Journal of
  Mathematical Biology}, vol.~79, no.~3, pp. 941--967, 2019.

\bibitem{Subramanian2020MICCAI}
S.~Subramanian, K.~Scheufele, N.~Himthani, and G.~Biros, ``Multiatlas
  calibration of biophysical brain tumor growth models with mass effect,'' in
  \emph{Medical Image Computing and Computer Assisted Intervention -- MICCAI
  2020}, A.~L. Martel, P.~Abolmaesumi, D.~Stoyanov, D.~Mateus, M.~A. Zuluaga,
  S.~K. Zhou, D.~Racoceanu, and L.~Joskowicz, Eds.\hskip 1em plus 0.5em minus
  0.4em\relax Cham: Springer International Publishing, 2020, pp. 551--560.

\bibitem{Subramanian20IP}
S.~{Subramanian}, K.~{Scheufele}, M.~{Mehl}, and G.~{Biros}, ``Where did the
  tumor start? an inverse solver with sparse localization for tumor growth
  models,'' \emph{Inverse Problems}, no.~36, 2020.

\bibitem{Scheufele2020}
K.~{Scheufele}, S.~{Subramanian}, and G.~{Biros}, ``Fully automatic calibration
  of tumor-growth models using a single mpmri scan,'' \emph{IEEE Transactions
  on Medical Imaging}, pp. 1--1, 2020.

\end{thebibliography}

\end{document}


\title{Supplementary: Ensemble inversion for brain tumor growth models with mass effect}
\author{Shashank Subramanian, Klaudius Scheufele, Naveen Himthani, Christos Davatzikos, \IEEEmembership{Fellow, IEEE}, and George Biros, \IEEEmembership{Member, IEEE}
\thanks{S. Subramanian, Oden Institute, University of Texas at Austin, Texas, USA e-mail: shashanksubramanian@{utexas}.edu.}
\thanks{K. Scheufele, Oden Institute, University of Texas at Austin, Texas, USA e-mail: klaudius.scheufele@{gmail}.com}
\thanks{N. Himthani, Oden Institute, University of Texas at Austin, Texas, USA e-mail: naveen@{oden}.utexas.edu}
\thanks{C. Davatzikos, University of Pennsylvania, Philadelphia, USA e-mail: christos.davatzikos@pennmedicine.{upenn}.edu}
\thanks{G. Biros, Oden Institute, University of Texas at Austin, Texas, USA e-mail: biros@{oden}.utexas.edu.}
}
\maketitle

\noindent \textbf{\textit{Model and inversion additional parameter values: }}
We report additional solver parameter values (based on~\cite{Subramanian:2019a}) in Tab.~\ref{tab:param_vals}. There is no subject-specific tuning of any parameter. The tumor initial condition (IC) reconstruction (\textbf{S}.1) is performed using a spatial resolution of $256^3$, to ensure smallest possible IC. The ensemble inversion for parameters (\textbf{S}.3) is performed in a downsampled resolution $160^3$ for faster solution times. We do not observe significant parameter inversion sensitivity between the two resolutions for the final step. If (\textbf{S}.3) is also performed in $256^3$, the full inversion time is about 4-6 hours on a V100 GPU. Our current inversion time is about 1-2 hours.
Finally, for the ensemble inversion, we observe that in rare cases some templates produce potentially unreliable solutions---we discard solutions with large condition number ($>1000$) of the finite difference 3x3 Hessian matrix (there are three scalar unknowns---$\kappa,\rho,\gamma$) of the objective function. 

\setlength\tabcolsep{5pt}
\begin{table}[!htbp]
    \caption{Model and inversion additional parameter values used in our simulations. Note that the  Lam\`e coefficients $\lambda$ and $\mu$ are determined by the Young's modulus and Poisson's ratio of the tissue-type.
    }\label{tab:param_vals}
    \centering
            \begin{scriptsize}
        \begin{tabular}{a|c}
            \hline
            Parameter			&			value \\
            \hline
            Young's modulus of (GM, WM, CSF, tumor)	(Pa)&	(2100, 2100, 100, 8000)  \\
             Poisson's ratio of (GM, WM, CSF, tumor) 	&	(0.4, 0.4, 0.1, 0.45)  \\
            Gaussian width, $\sigma$ (voxels)	& 1  \\
            Spacing between Gaussians $\delta$ (voxels) & 2 \\
            \hline
            Regularization parameter, $\beta$ & 1 \\
            Relative gradient tolerance		& 1E-3 \\
            Relative objective function change tolerance & 1E-3 \\
            Maximum number of quasi-Newton iterations & 50 \\
            \hline
    \end{tabular}
            \end{scriptsize}
\end{table}

\begin{figure}[!htbp]
    \centering
    \includegraphics[width=0.5\textwidth]{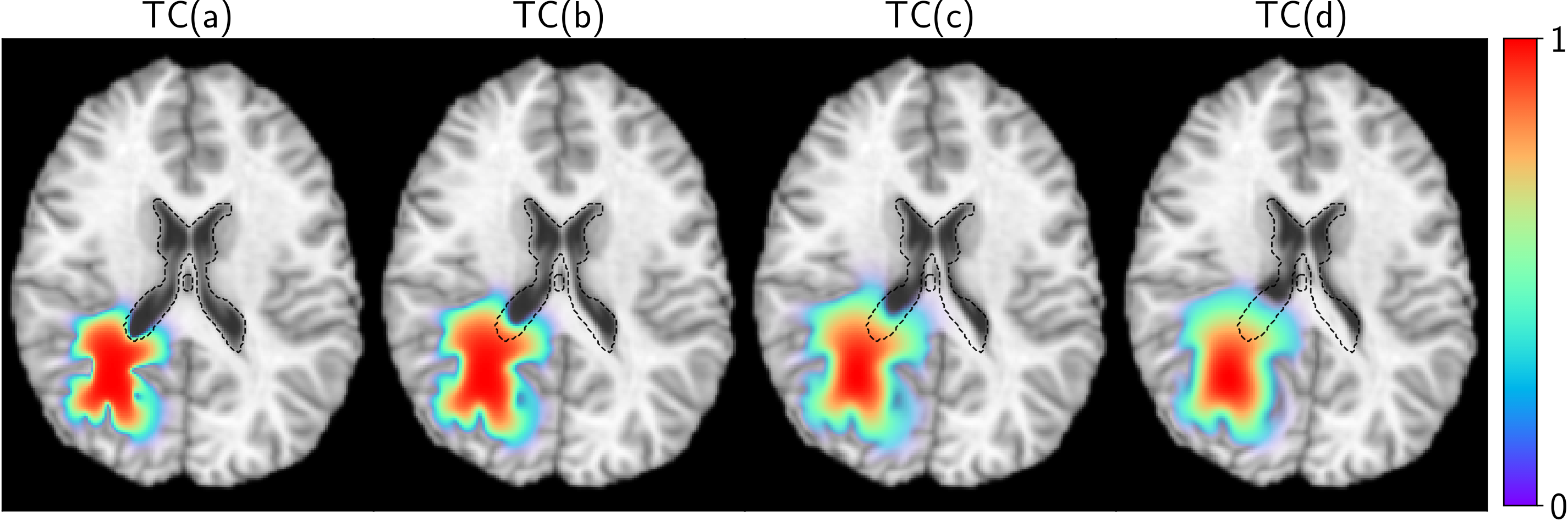}
    \caption{Synthetic data T1 MRIs generated with our forward model. The tumor concentration is overlaid (color) along with the undeformed ventricles (black dashed line) to indicate the variable extent of mass effect. The maximum displacements are approximately 0, 5.4, 8.9, and 13.3 mm, respectively.} \label{fig:syn}
\end{figure}

\begin{figure}[!htbp]
    \centering
    \includegraphics[width=0.45\textwidth]{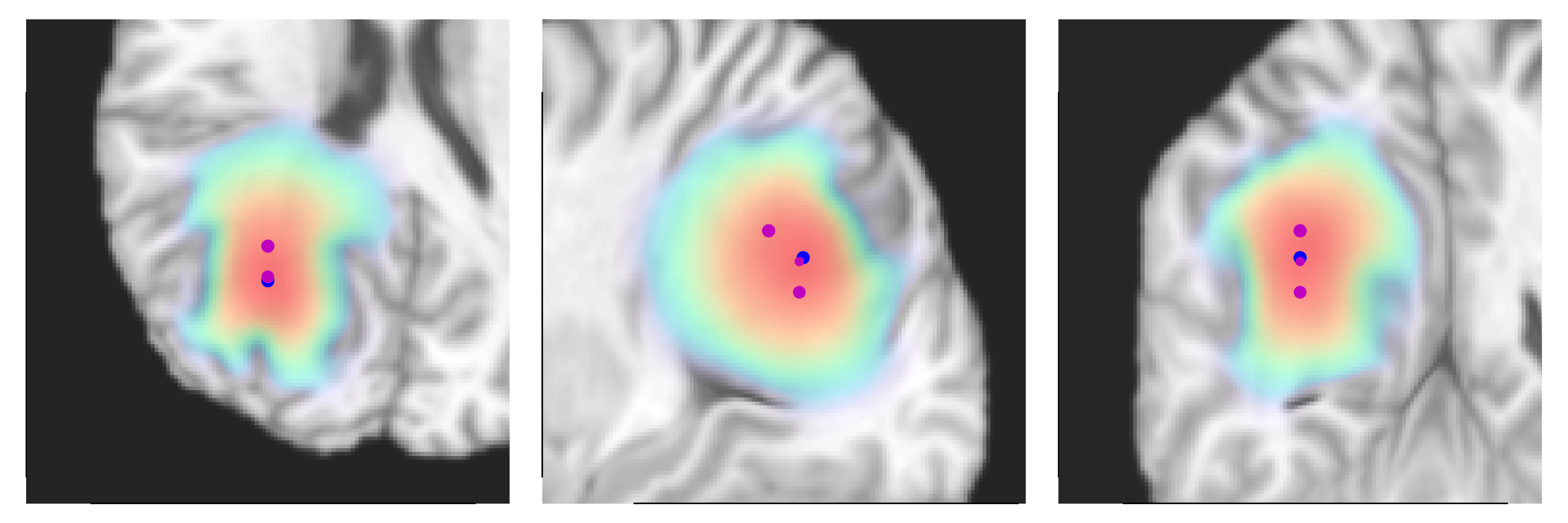}
    \caption{Estimated initial conditions $\vect{p}$ reconstructed using the algorithm in (\textbf{S}.1) for test-case TC(d). The size of the markers indicate the magnitude of activation. Further, the initiation locations are projected onto the respective 2D slice view (axial, sagittal, coronal) for ease of visualization. The blue marker is the ground truth and the red markers represent the reconstruction. The tumor concentration is overlaid (red: high, green: low). Due to ill-posedness, the exact reconstructions cannot be recovered.} \label{fig:syn-c0-recon}
\end{figure}

\medskip \noindent \textbf{\textit{Synthetic verification additional results: }}
We reproduce some of the solver verification results from our previous published work~\cite{Subramanian2020MICCAI}.
The synthetic data, i.e., the tumors along with the deformed $\vect{m}_0$ anatomy is shown in Fig.~\ref{fig:syn} for the four test-cases. 
We report the reconstruction errors for \textbf{Q1} (known $\vect{p}$ and $\vect{m}_0$) and \textbf{Q2} (known $\vect{m}_0$) in Tab.~\ref{tab:syn}. 
Reconstruction errors are less than 2\% for \textbf{Q1}. For \textbf{Q2}, the errors increase because we reconstructed $\vect{p}$ using the no-mass-effect model and the inverse problem is severly ill-posed (see~\cite{Subramanian20IP} for analysis of the non-convexity). But we can still recover the model parameters quite well: the mass effect (error in two-norm of the displacement norm $e_u$) is captured with relative errors less that 8\%; the reaction and diffusion coefficient also have good estimates of around 16\% and 17\% average relative errors, respectively.
We show the estimated tumor initial conditions for a specific test-case in Fig.~\ref{fig:syn-c0-recon} using (\textbf{S}.1). We observe that the reconstructed initial conditions are in a localized neighbourhood around the ground truth initial condition for all test-cases. Due to the inherent ill-posedness and model approximation errors, the exact initial locations are not recoverable.
\setlength{\tabcolsep}{6pt}
\begin{table*}[!htbp]
    \caption{Inversion results assuming \textbf{known} $\vect{m}_0$, the anatomy of the subject before tumor occurrence. 
    We report our reconstructions for parameters $\iota = \gamma, \rho, \kappa,$ and the maximum value of the displacement field two-norm (i.e., $u_\infty = \|u\|_\infty$ in mm). The ground truth parameters are indicated by $\iota^\star$ and the reconstructed values are $\iota$. We also report $e_\iota$, the relative error for parameters $\iota = \gamma,\rho,\kappa$; $e_u$ is the relative error in norm of the displacement field two-norm (i.e., $\|u\|_2$). 
    Finally, we report $\elltwoT$, the dice coefficient in reconstructing the tumor core.
The tumor initial condition used for inversion is indicated in the column ``Tumor IC" -- True IC indicates synthetic ground truth IC and Inverted IC indicates our reconstructions using (\textbf{S}.1) and (\textbf{S}.3).\label{tab:syn}
    (Note: when the ground truth ($^\star$) is zero, the errors are absolute errors; diffusion coefficient ($\kappa$) values are scaled by $\num{1E-2}$ for clarity).}
    \centering
            \begin{scriptsize}
    \makebox[\textwidth]{\centering
        \begin{tabular}{cc|cacaca|ca|a}
            \hline 
            Test-case   & Tumor IC & $(\gamma^\star,\gamma)$ & $e_\gamma$ & $(\rho^\star,\rho)$ & $e_\rho$ & $(\kappa^\star, \kappa)/0.01$ & $e_\kappa$ & $(u^\star_\infty, u_\infty)$ & $e_u$ & $\elltwoT$  \\
            \hline    
            {\textit{TC (a)}}& True IC & (0,0.0053) & \num{5.31E-3} & (12,12.01) & \num{9.58E-4} & (2.5,2.48) & \num{8.232e-03} & (0,0.07) & \num{1.92E-3} & \num{9.96E-1} \\
            {\textit{TC (b)}}& True IC & (0.4,0.4) & \num{1.58E-3} & (12,11.99) & \num{8.33E-5} & (2.5,2.49) & \num{1.412e-03} & (5.4,5.43) & \num{6.87E-3} & \num{9.96E-1} \\
            {\textit{TC (c)}}& True IC & (0.8,0.79) & \num{4.68E-4} & (10,10) & \num{6.3E-5} & (5,5) & \num{2E-5} & (8.98,8.97) & \num{1.12E-3} & \num{9.98E-1} \\
            {\textit{TC (d)}}& True IC & (1.2,1.19) & \num{6.67E-5} & (10,9.99) & \num{1E-5} & (2.5,2.5) & \num{1.456e-03} & (13.35,13.37) & \num{3.36E-3} & \num{9.97E-1} \\
            \hline
            {\textit{TC (a)}}& Inverted IC & (0,0.082) & \num{8.22E-2} & (12,9.82) & \num{1.81E-1} & (2.5,2.41) & \num{3.41E-2} & (0,0.11) & \num{8.25E-1} & \num{9.06E-1} \\
            {\textit{TC (b)}}& Inverted IC & (0.4,0.42) & \num{2.99E-2} & (12,10.13) & \num{1.55E-1} & (2.5,2.02) & \num{1.93E-1} & (5.4,5.18) & \num{2.17E-2} & \num{9.03E-1} \\
            {\textit{TC (c)}}& Inverted IC & (0.8,0.82) & \num{2.36E-2} & (10,8.62) & \num{1.38E-1} & (5,4.17) & \num{1.67E-1} & (8.98,8.55) & \num{3.54E-2} & \num{9.98E-1} \\
            {\textit{TC (d)}}& Inverted IC & (1.2,1.18) & \num{1.73E-2} & (10,8.5) & \num{1.49E-1} & (2.5,1.73) & \num{3.05E-1} & (13.35,12.25) & \num{7.68E-2} & \num{8.73E-1} \\
            \hline
            
    \end{tabular}}
           \end{scriptsize}
\end{table*}
\begin{figure*}[!htbp]
    \centering
    \includegraphics[width=\textwidth]{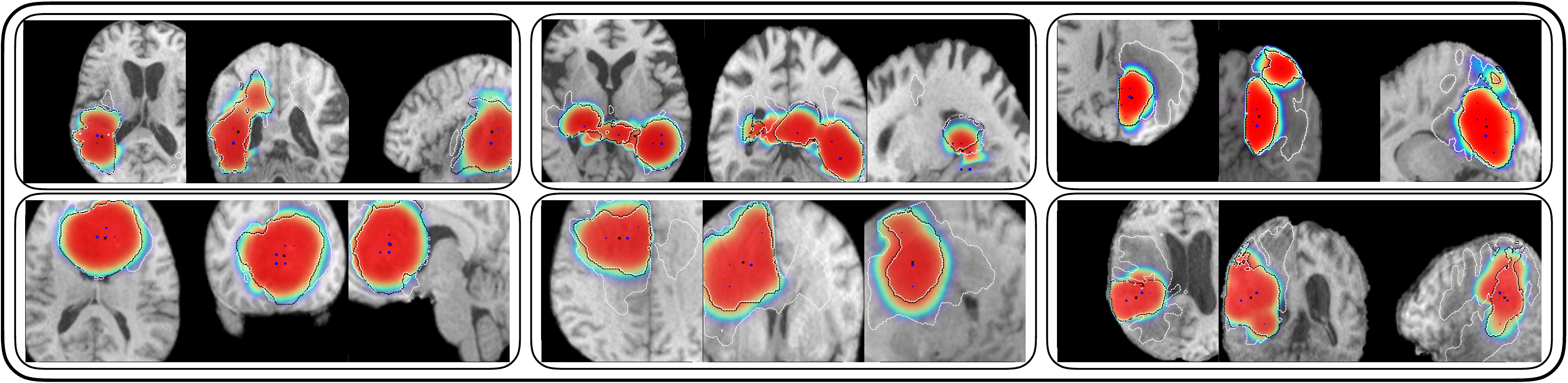}
    \caption{Estimated initial conditions $\vect{p}$ reconstructed using the algorithm in (\textbf{S}.1) for a few subjects. Each patient is a set of three images representing the axial, sagittal, and coronal 2D slices of the patient brain, zoomed into the tumor region.
    The blue markers represent the tumor initiation reconstruction and the black cross is the center of mass of the tumor core component. 
    The size of the markers indicate the magnitude of activation.
    Further, the initiation locations are projected onto the respective 2D slice view (axial, sagittal, coronal) for ease of visualization. 
    The tumor concentration is overlaid (red: high, green: low) with the black dashed line representing the tumor core data and the white dashed line representing edema. 
   } \label{fig:clin-c0-recon}
\end{figure*}

\begin{figure}[!htbp]
    \centering
    \includegraphics[width=0.48\textwidth]{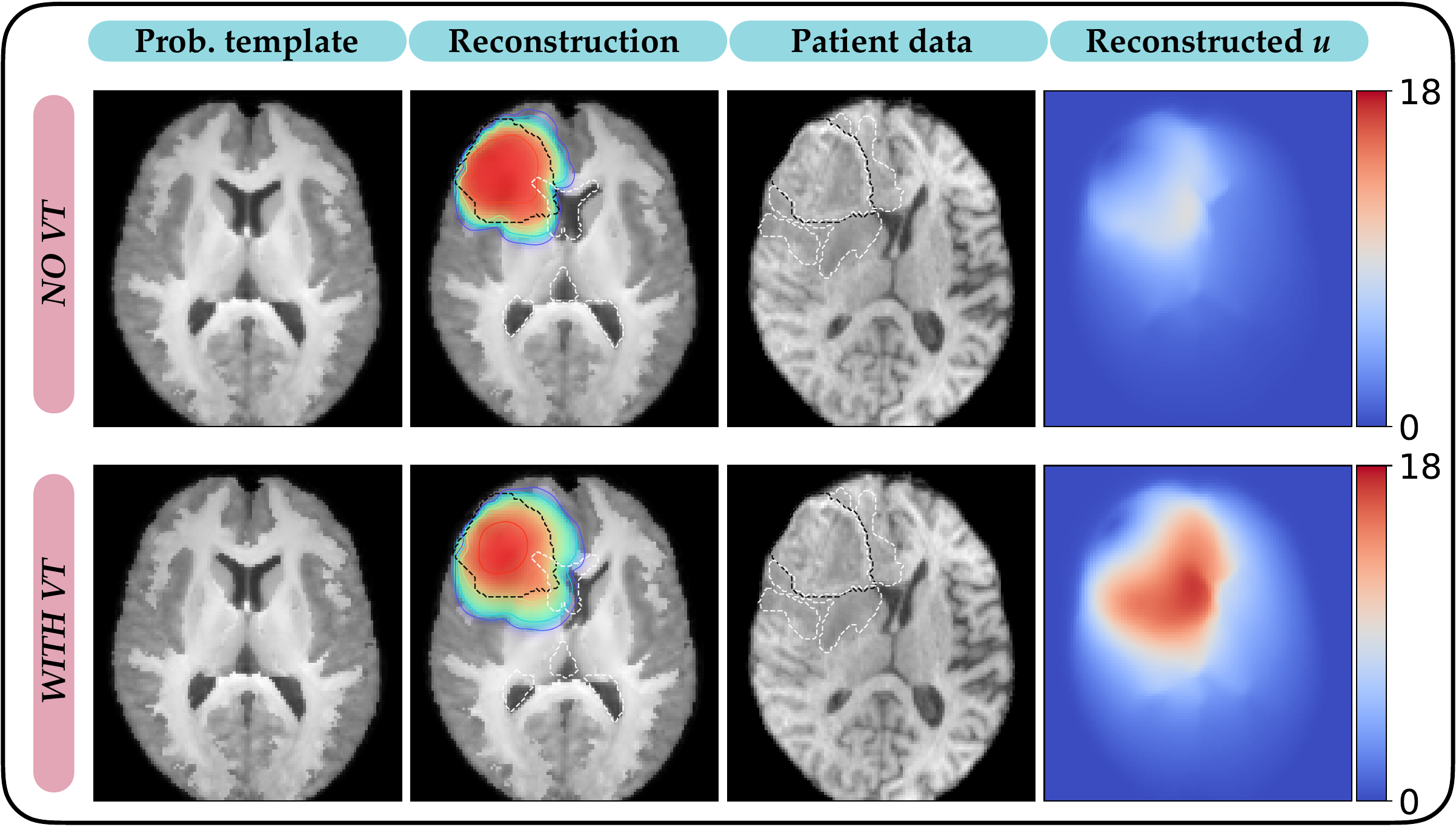}
    \caption{Reconstruction using two types of objective functions---with VT (VT misfit included) and no VT (only tumor misfit). The first column shows the averaged template T1 MRI, the second column shows the averaged patient reconstruction with the estimated tumor concentration (higher concentrations ($\sim$1) are indicated in red and lower ones in green) overlayed. We also show the underformed ventricle contour (white dashed line) to express the extent of mass effect, as well as the TC data contour (black dashed line). The third column shows the patient T1 MRI data with the TC data contour (black dashed line) and the edema data contour (white dashed line). The last column depicts the reconstructed average displacement (in mm) two-norm field.
We observe different reconstructions---the no VT result shows much smaller mass effect and deformation of ventricles and is qualitatively inaccurate.}\label{fig:VT}
\end{figure}

\medskip \noindent \textbf{\textit{Inversion additional results: }}
We provide some additional reconstruction details and experiments:
\begin{enumerate}[(i)]
    \item We show the initial condition reconstruction for a few GBM subjects (with multifocal examples) in Fig.~\ref{fig:clin-c0-recon}.
    The solver reconstructs sparse and localized tumor initial conditions (irrespective of the focality of the tumor data) across the clinical cohort. While certain initiation locations are close to the center of mass, we observe additional initiation sites are required to capture the complex hetereogeneous shape of brain tumors.
    For more qualitative results on clinical data and discussion, see our previous work~\cite{Scheufele2020}.
   \item We report our inversion results using two choices for number of templates (8 and 16) for precancer brain $\vect{m}_0$ approximation in Tab.~\ref{tab:clinical-sensitivity}. We observe insignificant sensitivity---the standard deviation of our reconstructed parameters and metrics are representative of the anatomical variation (and approximation error) between the patient and the chosen template.
   \item We observe that the inclusion of ventricles (VT) in the objective function (in the data misfit term $\| \vect{m}(1) - \vect{m}_p \|_2^2$) is important. While it does not make a significant difference for several patients, it does lead to different (and qualitatively inaccurate) results for certain patients. We compare our reconstructions in Fig.~\ref{fig:VT} and Tab.~\ref{tab:clinical-VT} for one such patient. We observe that the absence of the VT misfit term leads to erroneous reconstruction of mass effect (the average displacement norm is almost halved). This can be observed in our visualization as well as the reconstructed parameters and dice coefficient for VT. Another note: we do not include GM, WM, and CSF mismatches in our objective function (only tumor and VT included) since their segmentations (from ANTs) can be poor and they also show high anatomical dissimilarity.
\end{enumerate}
\setlength{\tabcolsep}{5pt}
\begin{table*}[!htbp]
    \caption{Inversion results sensitivity with number of templates for clinical data. 
    All results are repeated for two choices of number of templates---8 and 16.
 We report the mean ($\mu_\iota$) and standard deviation ($\sigma_\iota$) of the reconstructed parameters across the ensemble of templates; $\iota = \gamma, \rho, \kappa,$ and the maximum value of the displacement field two-norm (i.e., $u_\infty = \|u\|_\infty$ in mm). 
     We also report the probabilistic tumor core dice coefficient $\elltwoT$, the average runtime $\mu_T$, and standard deviation in runtime $\sigma_T$ in hh:mm:ss.
    (Note: the diffusion coefficient ($\kappa$) values are scaled by $\num{1E-2}$ for clarity; Patient AANN resulted in less than 8 templates for preselection due to the abnormally large ventricles---the inversion is identical for both template number choices).\label{tab:clinical-sensitivity}}
    \centering
            \begin{scriptsize}
        \begin{tabular}{ca|cccc|a|a}
            \hline
            Patient ID & Num. of templates & $(\mu_\gamma, \sigma_\gamma)$ & $(\mu_\rho, \sigma_\rho)$ & $(\mu_\kappa, \sigma_\kappa)/0.01$ & $(\mu_{u_\infty}, \sigma_{u_\infty})$ & $\elltwoT$ & $(\mu_T, \sigma_T)$ (in hh:mm:ss) \\
          \hline
          \textit{AAUJ} & 8 & (0.81, 0.065) & (9.83, 0.83) & (2.09,1.14) & (17.17, 1.05) & 0.871 & (01:13:51,00:18:27)\\
          \textit{AAUJ} & 16 & (0.82, 0.081) & (9.92, 0.75) & (1.87,0.89) & (17.57, 1.4) & 0.875 & (01:13:27,00:13:43)\\
            \hline
          \textit{AANN} & 8 & (0.59, 0.25) & (9.93, 0.61) & (2.28, 1.23) & (11.68, 4.91) & 0.853 & (01:27:24,00:06:23) \\
          \textit{AANN} & 16 & (0.59, 0.25) & (9.93, 0.61) & (2.28, 1.23) & (11.68, 4.91) & 0.853 & (01:27:24,00:06:23) \\
            \hline
          \textit{AARO} & 8 & (0.42, 0.109) & (10.54, 0.28) & (1.8, 0.37) & (7.42, 1.79) & 0.833 & (01:08:03,00:07:09) \\
          \textit{AARO} & 16 & (0.4, 0.089) & (10.64, 0.58) & (1.75, 0.38) & (7.23, 1.49) & 0.839 & (01:08:19,00:05:47) \\
            \hline
          \textit{AALU} & 8 & (0.51, 0.22) & (6.34, 0.45) & (0.97, 0.55) & (3.36, 1.36) & 0.928 & (01:06:52,00:07:55) \\
          \textit{AALU} & 16 & (0.36, 0.23) & (6.39, 0.52) & (1.1, 0.55) & (2.43, 1.48) & 0.942 & (01:05:22,00:08:14) \\
           \hline 
    \end{tabular}
           \end{scriptsize}
\end{table*}
\setlength{\tabcolsep}{5pt}
\begin{table*}[!htbp]
    \caption{ Inversion results for patient AAUJ by including ventricle data misfit (with VT) and excluding (no VT) it in the objective function.
 We report the mean ($\mu_\iota$) and standard deviation ($\sigma_\iota$) of the reconstructed parameters across the ensemble of templates; $\iota = \gamma, \rho, \kappa,$ and the maximum value of the displacement field two-norm (i.e., $u_\infty = \|u\|_\infty$ in mm). 
     We also report the probabilistic tumor core dice coefficient $\elltwoT$, the probabilistic ventricle dice coefficient $\elltwoVT$, the average runtime $\mu_T$, and standard deviation in runtime $\sigma_T$ in hh:mm:ss.
    \label{tab:clinical-VT}
    }
    \centering
            \begin{scriptsize}
        \begin{tabular}{ca|cccc|aa|a}
            \hline
            Patient ID & Objective function & $(\mu_\gamma, \sigma_\gamma)$ & $(\mu_\rho, \sigma_\rho)$ & $(\mu_\kappa, \sigma_\kappa)/0.01$ & $(\mu_{u_\infty}, \sigma_{u_\infty})$ & $\elltwoT$ & $\elltwoVT$ & $(\mu_T, \sigma_T)$ (in hh:mm:ss) \\

          \hline
          \textit{AAUJ} & no VT & (0.38, 0.19) & (10.99, 0.75) & (1.73,0.76) & (8.38, 4.08) & 0.926 & 0.461 & (01:08:45,00:06:04)\\
          \textit{AAUJ} & with VT & (0.82, 0.081) & (9.92, 0.75) & (1.87,0.89) & (17.57, 1.4) & 0.875 & 0.598 & (01:13:27,00:13:43)\\
           \hline 
    \end{tabular}
           \end{scriptsize}
\end{table*}
\begin{figure}[!htbp]
    \centering
    \includegraphics[width=0.5\textwidth]{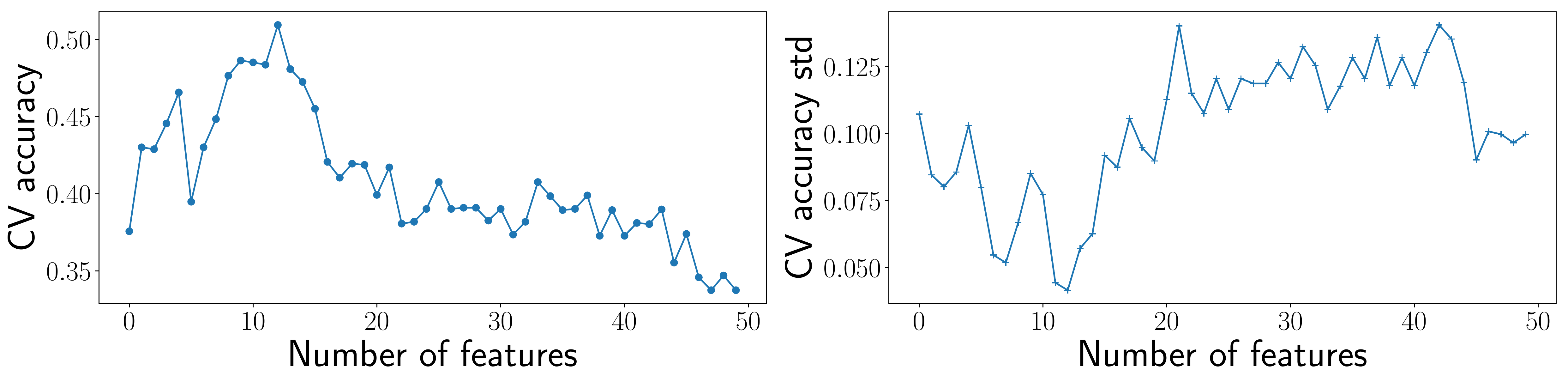}
    \caption{\textit{CV accuracy and standard deviation of accuracy values of a random forest classifier by sequentially adding features in descending order of their feature importances. We select the top 13 features due to their maximum CV accuracy and small accuracy std values.}}\label{fig:feature-selection}
\end{figure}

\medskip \noindent \textbf{\textit{Machine learning classifier details: }}
We employ a random forest classifier for our survival prediction task.
The hyperparameters of the random forest include maximum tree depth (taken as $3$) and number of estimators (taken as $100$), and are estimated during our cross-validation procedure.
For feature selection, we incrementally add features in their decreasing order of random forest feature importance and plot the cross-validated (CV) accuracy and standard deviation in accuracy in Fig.~\ref{fig:feature-selection}.  
The plot shows that the top 13 features provide the highest CV accuracy with low standard deviation in accuracy across the CV splits---these features are selected for survival classification.
Finally, we additionally experimented with other classifiers and regressors (logistic regression, linear regression, and elastic nets) but observed that \textbullet \, random forests provide the highest classification accuracy and \textbullet \, similar conclusions from the main text can be drawn from a regression analysis.

        \bibliographystyle{bib/IEEEtran}
\bibliography{bib/literature.bib}